\documentclass[aps,prb,10pt,twocolumn,amsmath,amssymb,longbibliography,superscriptaddress]{revtex4-2}

\pdfoutput=1

\usepackage[utf8]{inputenc}
\usepackage[english]{babel}
\usepackage[colorlinks=true,linkcolor=blue,citecolor=blue]{hyperref}
\usepackage{xcolor}
\usepackage{graphicx}
\usepackage[caption=false,margin=0pt]{subfig}

\begin{document}

\renewcommand{\vec}{\mathbf}
\renewcommand{\figurename}{FIG.}

\title{Electron-phonon coupling of Fe-adatom electron states on MgO/Ag(100)}

\author{Haritz Garai-Marin}
\affiliation{Physics Department, University of the Basque Country UPV/EHU, 48080 Bilbao, Basque Country, Spain}
\affiliation{Donostia International Physics Center (DIPC), Paseo Manuel de Lardizabal 4, 20018 Donostia-San Sebasti{\'a}n, Spain}

\author{Julen Ibañez-Azpiroz}
\affiliation{Centro de F{\'i}sica de Materiales, Universidad del Pa{\'i}s Vasco UPV/EHU, 20018 San Sebasti{\'a}n, Spain}
\affiliation{IKERBASQUE Basque Foundation for Science, 48013 Bilbao, Spain}

\author{Peio Garcia-Goiricelaya}
\affiliation{Centro de F{\'i}sica de Materiales, Universidad del Pa{\'i}s Vasco UPV/EHU, 20018 San Sebasti{\'a}n, Spain}

\author{Idoia G. Gurtubay}
\author{Asier Eiguren}
\affiliation{Physics Department, University of the Basque Country UPV/EHU, 48080 Bilbao, Basque Country, Spain}
\affiliation{Donostia International Physics Center (DIPC), Paseo Manuel de Lardizabal 4, 20018 Donostia-San Sebasti{\'a}n, Spain}

\date{\today}

\begin{abstract}
	We study the strength of the electron-phonon interaction on Fe single adatoms on MgO/Ag(100) based on many-body \textit{ab-initio} spin collinear calculations. In particular, we analyze the relative importance of the substrate and, among other results, we conclude that the interface electron state of Ag(100) plays a prominent role in determining the electron-phonon coupling of localized Fe electron states. The analysis of the hybridization of the adatom with the substrate reveals qualitative differences for even or odd coverages of MgO, affecting significantly the spectral structure and strength of the electron-phonon coupling. Our calculations indicate that the electron-phonon interaction is very strong for $\le$~1 layers of MgO, while it is sharply suppressed for larger coverages, a trend that is consistent with recent experimental findings.
\end{abstract}

\maketitle

\section{Introduction}

Controlling the magnetic moment of individual atoms and understanding the underlying physics is an extraordinary technological and scientific challenge due to its potential applications in high-density storage devices~\cite{Natterer2017} as well as for quantum computing~\cite{Khajetoorians2011a}, among other cutting edge research fields. Their magnetic structure has been intensively studied by theoretical density-functional theory (DFT)~\cite{Lang1994,Lounis2010,Ibanez-Azpiroz2017a}, model spin Hamiltonians~\cite{Lorente2009,Delgado2017}, and more recently by \textit{ab-initio} based multiplet calculations~\cite{Wolf2020}. In addition, inelastic electron tunneling spectroscopy~\cite{Heinrich2004,Hirjibehedin2006,Hirjibehedin2007,Khajetoorians2011,Donati2013,Hermenau2018}, X-ray magnetic circular dichroism~\cite{Donati2014,Donati2014a,Donati2016,Baltic2018}, spin-polarized scanning tunneling microscope~\cite{Meier2008,Loth2010,Loth2010b,Paul2017,Natterer2018} and electron paramagnetic resonance~\cite{Baumann2015b,Natterer2017} experiments have further characterized the spin-dynamics, showing that it is possible to create long-living magnetic quantum states in individual adatoms.

The state-of-the-art research of the field is currently focused on understanding the physical mechanisms involved on the spin dynamics, with the ultimate goal of achieving the control and stability of single atoms with long spin relaxation times like those found for Fe and Ho on MgO/Ag(100)~\cite{Paul2017,Donati2016}. The magnetic stability is determined by the strength of the energy barrier separating the ground state and the excited magnetic states. At low temperatures, the main relaxation mechanisms of the magnetic moment are associated to quantum tunneling of magnetization~\cite{Gatteschi2003} and the interactions with the environment~\cite{Delgado2017}, where it is expected that the main contribution is due to the coupling with the electron and phonon bath of the system. In order to stabilize the magnetic moment of the adatoms, insulating decoupling layers such as Cu\textsubscript{2}N or MgO have been used successfully~\cite{Hirjibehedin2007,Donati2016}. The insulating layers effectively reduce the electronic hybridization between the adatom and the conducting substrate, as thoroughly analyzed in the literature~\cite{Delgado2017,Ibanez-Azpiroz2017,Shehada2021}. However, the role of substrate phonons has received far less attention to date, despite clear indications of their importance in the relaxation mechanism~\cite{Donati2020}.

The so called electron-phonon interaction is essentially the effect of phonons on the electronic structure. The matrix elements become relevant when studying this coupling perturbatively, which are given by,
\begin{equation}\label{eq:matrix-element1}
	g_{if}^{\eta,ss'}
	=
	\left< \psi_i^{s} \left| \dfrac{\partial V^{ss'}}{\partial \xi^\eta} \right| \psi_f ^{s'}\right>
.\end{equation}
Above, $\partial V^{ss'}/\partial \xi^\eta$ represents the potential induced by a phonon mode $\eta$ with displacement vector $\xi^\eta$, while $\psi^{s}_i$ and $\psi^{s'}_f$ respectively represent the initial and final electron states with spin components $s$ and $s'$. $g_{if}^{\eta,ss'}$ can be naturally divided into two pieces; the spin-diagonal ($s=s'$) and spin-flip ($s\neq s'$) parts. The former is mostly driven by the Hartree term and is therefore commonly regarded as the dominant contribution in most materials~\cite{Giustino2017}. The spin-flip part, in turn, originates from the relativistic spin-orbit interaction that scales as $v^2/c^2$, with $v$ the electron velocity and $c$ the speed of light; among other properties, the spin-flip process determines the phonon-driven spin-lifetime of electron states~\cite{Roychoudhury2018,Lunghi2019}. Given their different nature, the spin-diagonal and spin-flip terms are expected to show marked differences in magnetic adatoms, \textit{e.g.}, in their total strength as well as dependence on the insulating coverage. To our knowledge, neither of these terms has been computed by \textit{ab-initio} means in single adatoms.

In this paper we focus on the spin-diagonal electron-phonon contribution and present \textit{ab-initio} calculations on a Fe adatom deposited on the MgO/Ag(100) surface. Our analysis reveals that the special features of the adsorbing surface, such as interface states, are of great importance on the electron-phonon scattering processes. Additionally, we study the dependence of the electron-phonon coupling strength on the number of MgO layers. We clarify the role of hybridization between the adatom and the substrate for different coverages, finding that the effect of the electron-phonon interaction is strongly suppressed for two or more layers of MgO, a trend that is consistent with recent experimental findings~\cite{Paul2017}.

The paper is organized as follows. In Section~\ref{sec:computational-details} we summarize the technical details of the formalism used to perform the DFT calculations, and we explain the computational method used to compute the vibrational structure and the electron-phonon matrix elements. Section~\ref{sec:results} contains the calculations and analysis of the effect of the spin-diagonal contribution to the electron-phonon interaction on the Fe adatom adsorbed on the MgO/Ag(100) surface. In particular, we introduce the electronic and vibrational properties of the MgO/Ag(100) surface and the adatom (Sec.~\ref{sec:electronic-and-vibrational-properties}), and we follow with the calculations and analysis of the electron-phonon coupling (Sec.~\ref{sec:electron-phonon-coupling}). Finally, Section~\ref{sec:conclusions} summarizes the main results and conclusions.

\section{Computational details}\label{sec:computational-details}

The first principles calculations were preformed using the DFT~\cite{Hohenberg1964,Kohn1965} formalism implemented in the SIESTA code~\cite{Soler2002}, based on numerical linear combination of atomic orbitals (LCAO) basis sets and pseudopotentials (PPs)~\cite{Pickett1989}. Optimized basis sets are used for silver and oxygen, a triple-zeta plus 2 polarization orbitals for magnesium and a triple-zeta plus 3 polarization orbitals for iron. Atomic cores are represented using separable~\cite{Kleinman1982} norm-conserving PPs~\cite{Hamann1979}. The generalized gradient approximation parametrized by Perdew, Burke and Ernzerhof~\cite{Perdew1996} (PBE-GGA) has been used for the exchange-correlation functional. The spin polarized calculations were done using the collinear formalism.

Calculations of two different systems are presented in this paper. On the one hand, the clean MgO/Ag(100) surface, which is modeled using a slab system consisting of eleven silver atoms with a monolayer of magnesium oxide on both terminations. Coverages of MgO from 0 to 4 monolayers (MLs) are considered, with a minimum vacuum region of 18 {\AA} to prevent interaction between slabs for 1~ML coverage. We also have considered the free standing MgO surface on our calculations to disentangle the role of electronic interactions on the properties of the adatom. In addition, we also consider a $4\times4\times1$ super-cell of the original MgO/Ag(100) surface, with an iron adatom on top of an oxygen site. Integrations over the Brillouin zone are done using a $4\times4\times1$ Monkhorst-Pack mesh~\cite{Monkhorst1976} for the clean MgO/Ag(100) slab, and the $\Gamma$ point for the super-cell with the iron adatom. The occupations are calculated by the Fermi-Dirac distribution with an electronic temperature of 300 K to accelerate selfconsistency. Additionally we have used a mesh cutoff of 600 Ry for our calculations, together with the \texttt{Grid.CellSampling} parameter on the adatom system to mitigate the egg-box effect on atomic forces and properly determine the soft modes of the adatom.

The usual approach to compute the lattice dynamics of a system is to calculate the dynamical matrix of the whole system. This can be done using Density Functional Perturbation Theory (DFPT)~\cite{Baroni2001}, obtaining the dynamical matrix for the desired phonon momentum, or by using the so-called direct method~\cite{Frank1995,Parlinski1997}, obtaining the interatomic force constants, which can be used to compute the dynamical matrix. DFPT is a robust, efficient and accurate method to compute the lattice dynamics of bulk systems, surfaces and small molecules~\cite{Baroni2001}, but it becomes impractical in a super-cell structure with hundreds of atoms. For this reason, we have instead employed the direct method to compute the lattice dynamics of the system. In this method, each atom is displaced in every cartesian direction, and the interatomic force constants are obtained differentiating the Hellmann-Feynman forces, where the derivative is approximated by a centered finite difference formula,
\begin{equation}
	\Phi^{\alpha,\beta}_{\kappa,\tau}
	=
	-\dfrac{F^\alpha_\kappa(\vec{R}+\vec{u}^\beta_\tau) - F^\alpha_\kappa(\vec{R}-\vec{u}^\beta_\tau)}{2|\vec{u}^\beta_\tau|}
.\end{equation}
Here $\vec{u}^\beta_\tau$ and $F^\alpha_\kappa$ are the displacement vector of the atom $\tau$ along direction $\beta$ and the Hellmann-Feynman force of atom $\kappa$ in the direction $\alpha$, respectively. The potential induced by the phonons, which is required to calculate the electron-phonon matrix elements, is computed by differentiating the Kohn-Sham potential ($V_\mathrm{KS}$), also approximated by centered finite differences,
\begin{equation}
	\dfrac{\partial V^{ss'}}{\partial u^\beta_\tau}
	=
	\dfrac{ V^{ss'}_\mathrm{KS}(\vec{R}+\vec{u}^\beta_\tau) - V^{ss'}_\mathrm{KS}(\vec{R}-\vec{u}^\beta_\tau) }{2|\vec{u}^\beta_\tau|}
.\end{equation}
The potential induced by a phonon $\eta$ that appears in Eq.~\eqref{eq:matrix-element1} is obtained by a linear combination of $\partial V^{ss'}/\partial u^\beta_\tau$.

In general, in a system with $N$ atoms, a naive implementation of this method would require $3\times N$ standard DFT calculations. The number of displacements to be considered can be greatly reduced by making use of the symmetries of the system and displacing only the non-equivalent atoms along the non-equivalent directions~\cite{Parlinski1997}. As an example, the usage of symmetries enables to compute the complete interatomic force constants of the $4\times4\times1$ 1~ML MgO/Ag(100) super-cell, which has P4/mmm space group symmetry, calculating the Helmann-Feynman forces for only 16 different configurations, recovering the rest using symmetry operations (See Ref.~\citenum{Parlinski1997} for more information).

Adding the iron adatom to the calculation breaks the translational and inversion symmetry of the system, while the 4-fold symmetry is still maintained. This means that one would be forced to consider hundreds of different self-consistent calculations if the translational symmetry of the clean substrate was not exploited. As adding the adatom changes only the neighboring interatomic force constants and leaves the substrate unaltered, in this paper we reduce the number of calculations by considering the interatomic force constants of the MgO/Ag(100) substrate without the adatom as an starting point. Then, we include the local modifications of the interatomic force constants due to the adatom by computing the interatomic force constants connected to the adatom and the oxygen underneath it. Thereby, the forces of only 22 \footnote{16 for the substrate + 3 for the iron + 3 for the oxygen below it.} displacements are needed to completely determine the vibrational modes of the Fe adatom on a $4\times4\times1$ 1~ML MgO/Ag(100) super-cell. More importantly, given that $\partial V^\beta_\tau$ is generally well localized around the displaced atom, we also can safely compute the induced potentials following this procedure.

While there are available programs to compute the interatomic force constants using the symmetries of the system, to our knowledge none of them has the capability of calculating the potential induced by the displacement of the atom, which is essential to compute the electron-phonon matrix elements. For this reason, we have implemented the method following the formulas found on Ref.~\citenum{Chaput2011} for obtaining the interatomic force constants, and obtained the induced potentials using symmetry operations. Additionally, we also compute the dynamical matrix from the interatomic force constants, which is diagonalized to obtain the energies $\omega_\nu$ and polarization vectors $\xi^\nu$ of the vibrational modes.

The spin-diagonal electron-phonon matrix elements are computed as
\begin{equation}\label{eq:matrix-element2}
	g^{\eta}_{if}
	=
	\sum_{ss'}
	g_{if}^{\eta,ss'} \delta_{ss'}
,\end{equation}
where $g_{\eta,if}^{ss'}$ is defined in Eq.~\eqref{eq:matrix-element1}. We compute the electron-phonon matrix elements in Fourier space, in order to take advantage of the computational scheme already developed by our group (See Refs.~\citenum{Garcia-Goiricelaya2018,Garcia-Goiricelaya2019,Garcia-Goiricelaya2020,Lafuente-Bartolome2020,Lafuente-Bartolome2020a} for most recent activity). To this end, we transform the wave-functions given by SIESTA in an atomic basis set ($l,m$) to a plane wave basis ($\vec{G}$) using Eqs. 23 and 24 from Ref.~\cite{Soler2002}:
\begin{equation}\label{eq:Fourier-transform}
	\psi(\vec{G})
	=
	\sum_{l=0}^{l_{max}} \sum_{m=-l}^{l}
	\psi_{lm}(G)Y_{lm}(\vec{\hat{G}})
,\end{equation}
and
\begin{equation}\label{eq:radial-transform}
	\psi_{lm}(G)
	=
	(-i)^l \int_{0}^{\infty}
	r^2 j_{l}(Gr)\psi_{lm}(r) dr
.\end{equation}
Here  $Y_{lm}$ and $j_l$ are spherical harmonics and spherical Bessel functions, respectively.

\section{Results}\label{sec:results}

\subsection{Electronic and vibrational properties}\label{sec:electronic-and-vibrational-properties}

In this section we focus on describing ground state properties. We have chosen the Ag(100) surface covered with three MgO layers as the reference system analyzed in Secs.~\ref{subsec:clean-substrate} and \ref{subsec:fe-adatom-on-mgoag100}, given that it best exemplifies the central features of interest. The analysis of the MgO layer dependence is done in Sec.~\ref{sec:layer-dependent-electronic-and-vibrational-properties}.

\subsubsection{Clean substrate}\label{subsec:clean-substrate}

\begin{figure}[tbph!]
	\centering
	\captionsetup[subfigure]{labelformat=empty}
	\subfloat[\label{fig:bands-slab_a}]{}
	\subfloat[\label{fig:bands-slab_b}]{}
	\includegraphics[width=\linewidth]{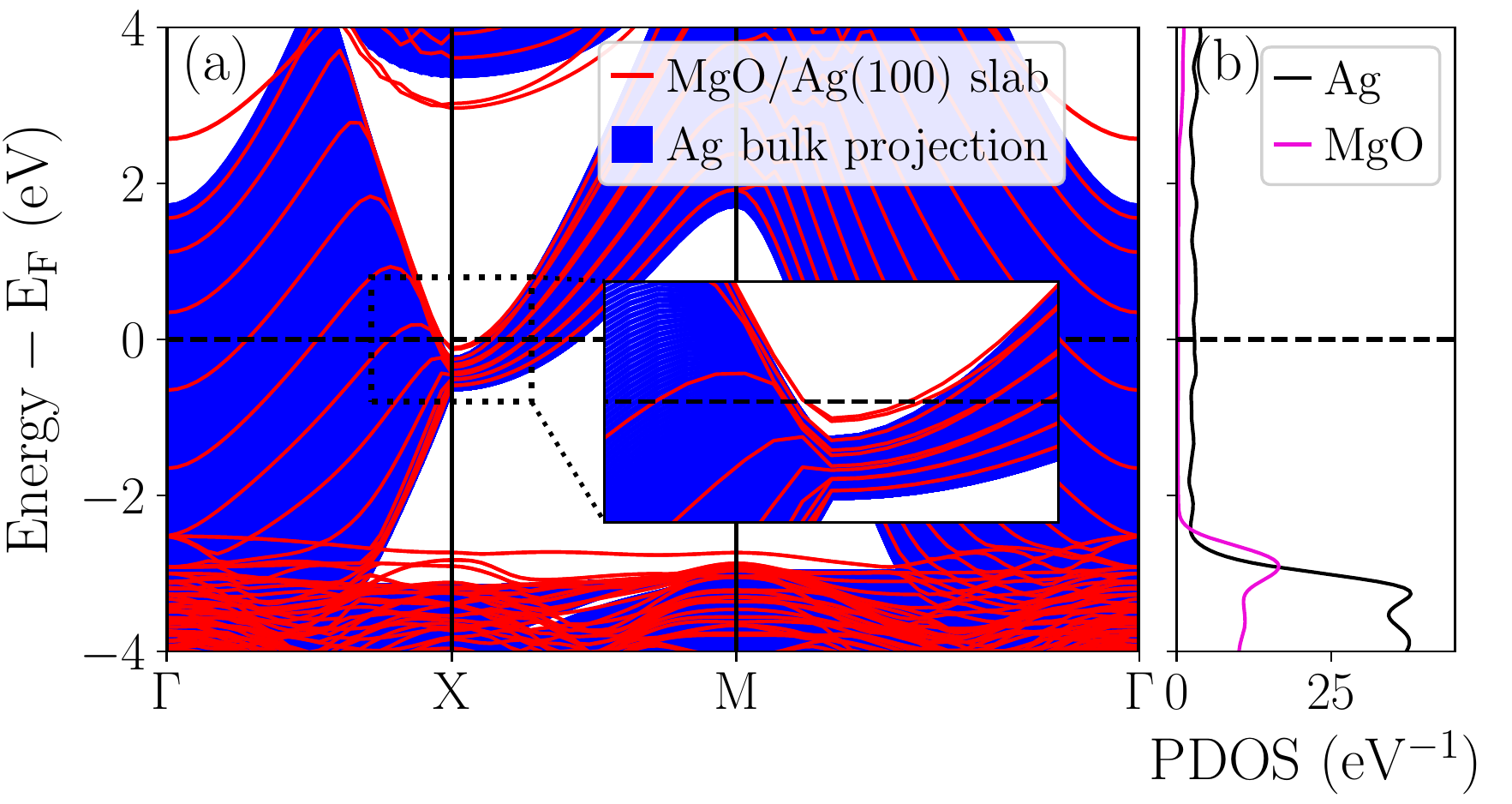}
	\vspace{-2\baselineskip}
	\caption{\label{fig:bands-slab}(a) The calculated band structure of the 3~ML MgO/Ag(100) slab (red lines) together with the bulk band projection of silver (solid blue). The inset shows the detail close to the Fermi energy of the interface state at the high symmetry $X$ point. (b) The calculated projected DOS of the MgO layer and the Ag(100) substrate. The dashed line represents the Fermi level.}
\end{figure}

\begin{figure}[b!]
	\centering
	\includegraphics[width=\linewidth]{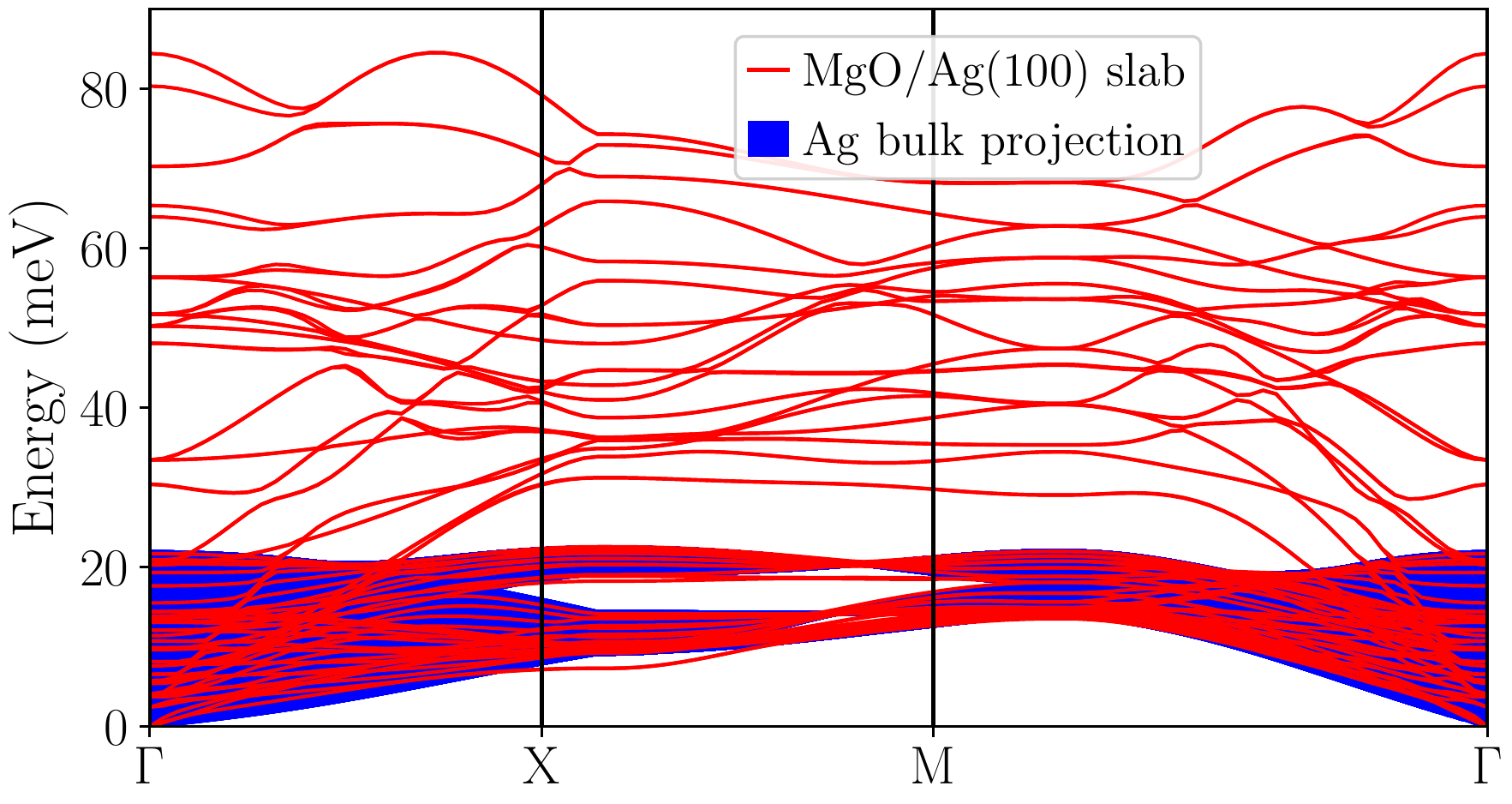}
	\caption{\label{fig:phonons-slab_3ML} Calculated phonon dispersion of the 3~ML MgO/Ag(100) slab (red lines) and silver bulk phonon projection (solid blue).}
\end{figure}
Figure~\ref{fig:bands-slab} shows the electronic band structure of the 3~ML MgO/Ag(100) surface. The projected density of states (PDOS) is shown in \figurename~\ref{fig:bands-slab_b}, where the electronic states of the MgO layer are below 2~eV from the Fermi level. Indeed, the meV energy window of phonon energies around the Fermi level is exclusively dominated by silver states. Focusing on the band structure in \figurename~\ref{fig:bands-slab_a}, the blue area represents the bulk band projection of the silver (100) surface. It is known from previous works that the Ag(100) surface hosts a surface state close to the Fermi energy at the $X$ point of the surface Brillouin zone~\cite{Kolb1981,Reihl1984,Altmann1986,Reihl1987,Erschbaumer1991,Savio2001}. Our calculation shows an electronic state localized on the MgO/Ag(100) interface with similar energy around the same region. As we will show, this interface state will play an important role in the conduction properties and on the electron-phonon coupling due to its energy and proximity to the surface.

In \figurename~\ref{fig:phonons-slab_3ML} we present the calculated phonon dispersion of the 3~ML MgO/Ag(100) surface. The blue area represents the bulk phonon projection of the Ag(100) surface and the red lines show the calculated phonon dispersion of the MgO/Ag(100) slab, which is very similar to that of the clean Ag(100) surface at least up to 20~meV~\cite{Heid2003}. The higher energy modes correspond to vibrations of MgO. The first acoustic mode of the MgO layer is completely mixed with silver oscillations in the 10-20~meV energy range.

\subsubsection{Fe adatom on MgO/Ag(100)}\label{subsec:fe-adatom-on-mgoag100}

We analyze now the properties of the Fe adatom deposited on the $4\times4\times1$ 3~ML-MgO/Ag(100) super-cell. We found the oxygen top adsorption site to be energetically the most favorable one for iron, in agreement with previous studies~\cite{Baumann2015b,Baumann2015,Paul2017}. After relaxation, the oxygen atom is slightly displaced upwards with a resulting Fe-O bond of 2.01~{\AA}, in agreement with Ref.~\citenum{Baumann2015}. The system develops a magnetic ground state, with a local magnetic moment of 3.97~$\mu_B$ for the iron adatom, and 0.1~$\mu_B$ spread over the surrounding atoms.

\begin{figure}[tbph!]
	\centering
	\includegraphics[width=\linewidth]{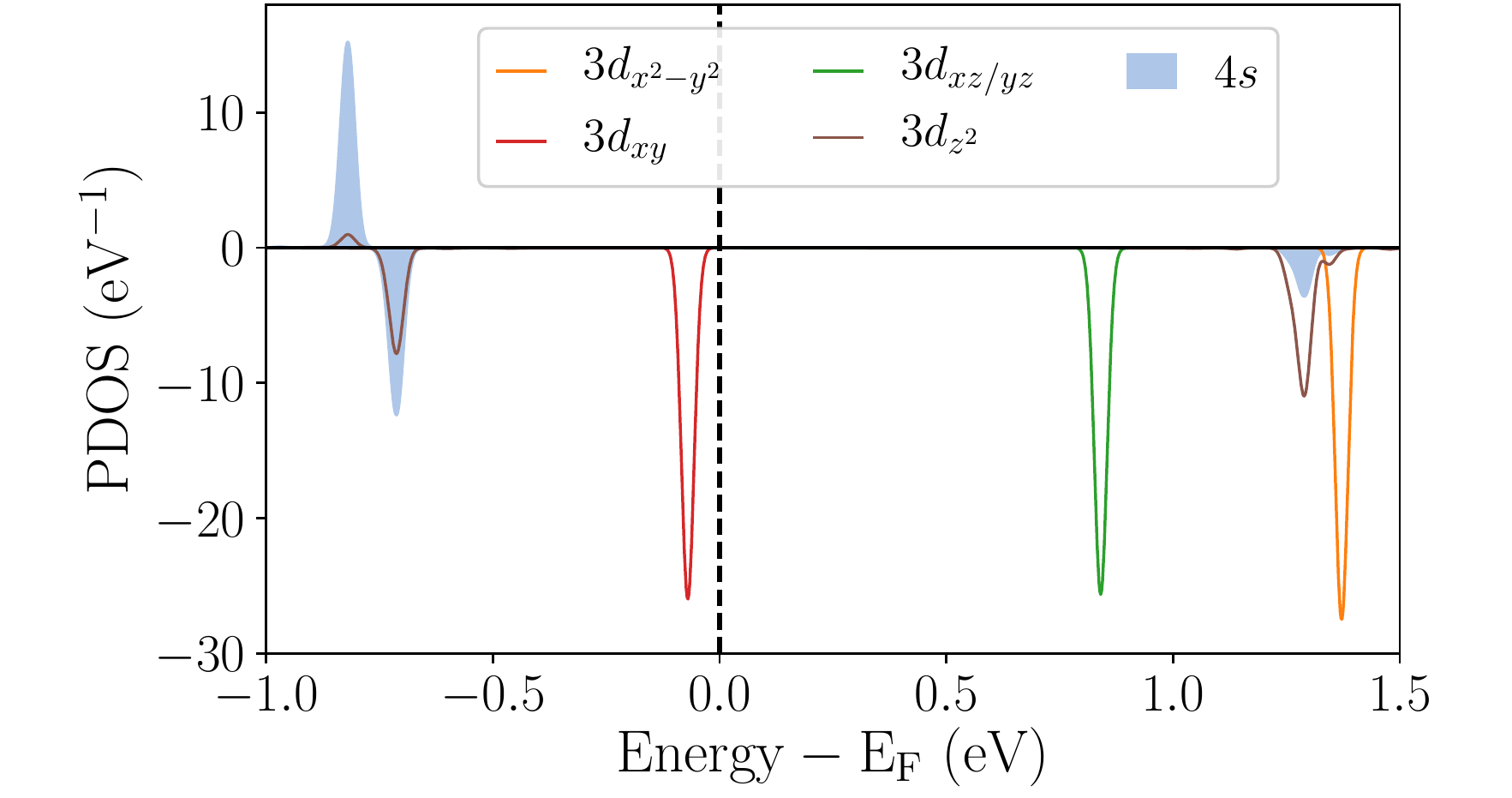}
	\caption{\label{fig:pdosfe} Projected DOS of iron's $3d$ (solid line) and $4s$ (solid background) orbitals for Fe on 3~ML MgO/Ag(100). Positive and negative values of the PDOS indicate majority and minority spin channels, respectively. The Fermi energy is marked by a vertical dashed line.}
\end{figure}

In \figurename~\ref{fig:pdosfe} we show the spin resolved PDOS projected on the Fe adatom. The occupation of the valence $3d$ electrons has a similar configuration to the free atom; the five majority states are filled, whereas only one minority orbital is fully occupied; note that the $3d_{z^2}$ is partially occupied because it is highly hybridized with the $4s$ orbital. The only orbital that falls in the meV range from the Fermi level is the minority $3d_{xy}$, which will therefore play a central role in the forthcoming analysis of the electron-phonon interaction.

\begin{figure}[tbph!]
	\centering
	\includegraphics[width=\linewidth]{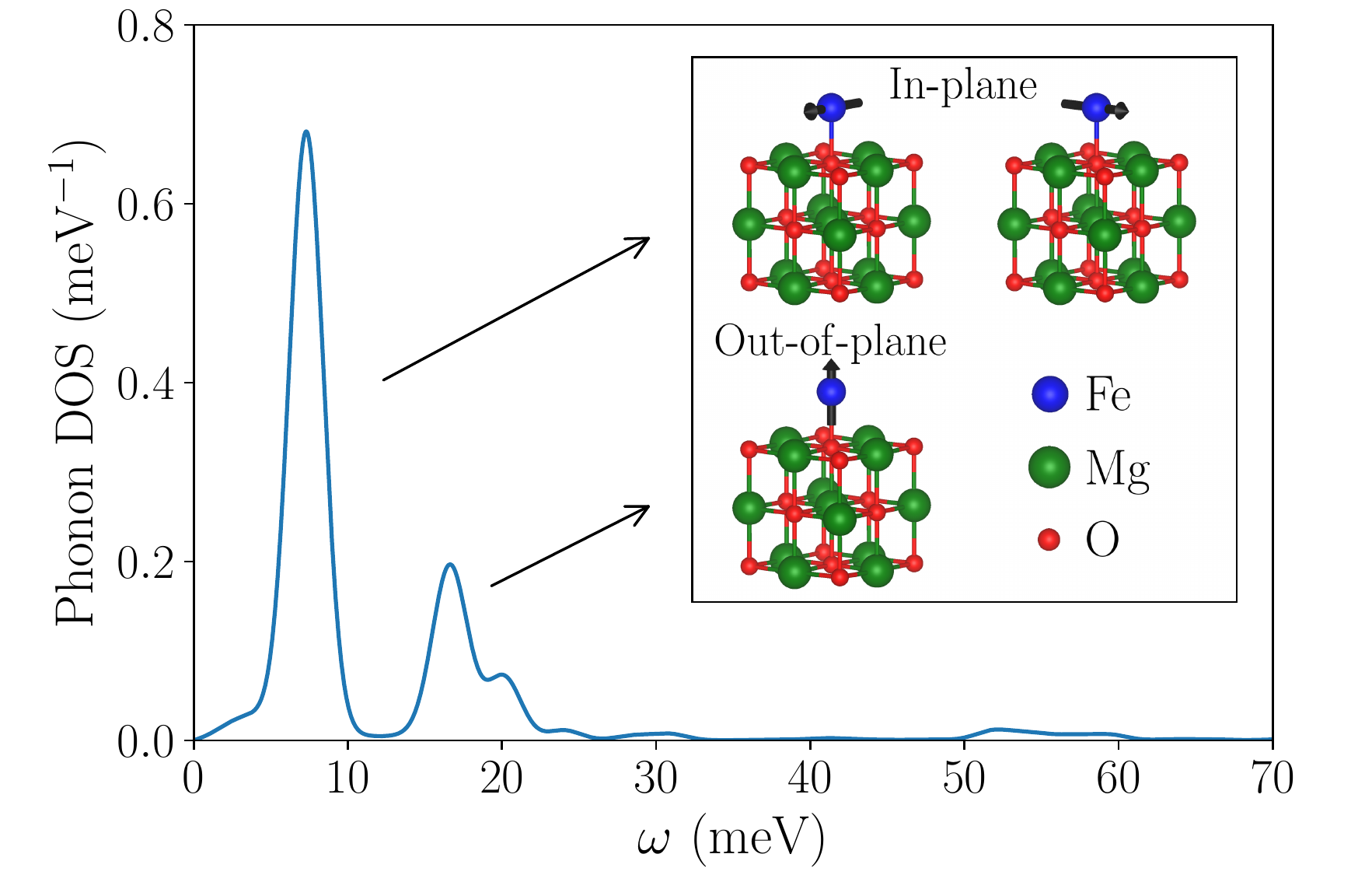}
	\caption{\label{fig:phdosfe} Phonon DOS projected onto the Fe adatom deposited on 3~ML MgO/Ag(100). The inset shows the MgO layers and the Fe adatom, with the arrows indicating the polarization vector of the vibrational modes localized on the adatom.}
\end{figure}

Next we analyze the vibration modes of the 3~ML Fe-MgO/Ag(100) system. The phonon DOS projected on the iron adatom is shown in \figurename~\ref{fig:phdosfe}. The figure reveals a vibrational mode completely localized in iron at 7.3~meV. Physically, this peak corresponds to two degenerate modes of the adatom oscillating parallel to the surface. At higher energies, in the 12-25~meV range, we observe the presence of multiple modes with an out-of-plane polarization and with a more pronounced mixing with the substrate, as inferred from the broadening of the peak.

\subsubsection{Layer dependent electronic and vibrational properties}\label{sec:layer-dependent-electronic-and-vibrational-properties}

\begin{figure}[tbph!]
	\centering
	\captionsetup[subfigure]{labelformat=empty}
	\subfloat[\label{fig:adsorption_1ML}]{}
	\subfloat[\label{fig:adsorption_2ML}]{}
	\includegraphics[width=\linewidth]{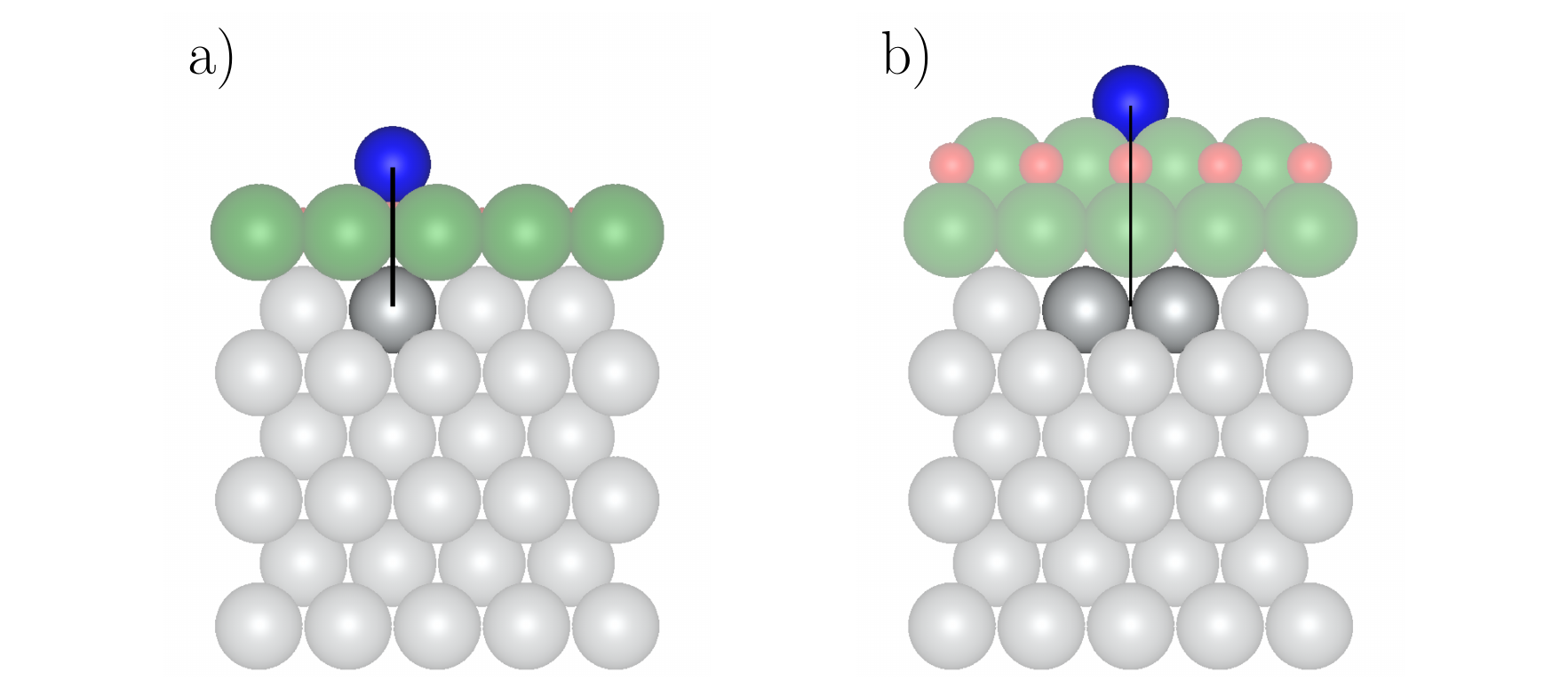}
	\vspace{-2\baselineskip}
	\caption{\label{fig:adsorption} Positioning of the iron adatom respect to the Ag(100) substrate for an (a) odd and (b) even number of MgO layers. The vertical black line is a guide for the eye.}
\end{figure}

\begin{figure*}[t!]
	\centering
	\captionsetup[subfigure]{labelformat=empty}
	\subfloat[\label{fig:isosurface_1ML}]{}
	\subfloat[\label{fig:isosurface_2ML}]{}
	\subfloat[\label{fig:isosurface_3ML}]{}
	\includegraphics[width=\linewidth]{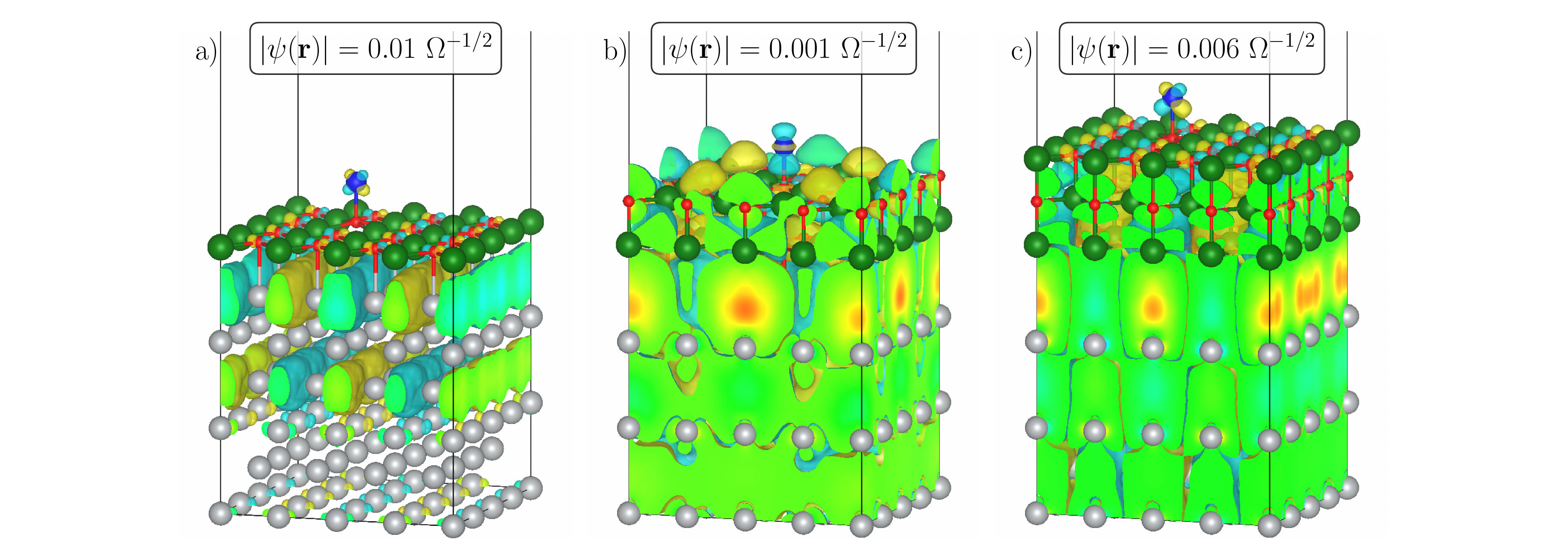}
	\vspace{-2\baselineskip}
	\caption{\label{fig:isosurface} Isosurface of the interface state discussed in \figurename~\ref{fig:bands-slab} for different coverages of MgO: a) 1~ML, b) 2~ML and c) 3~ML. The value of the isosurface has been chosen to make visible the hybridization with the iron adatom. $\Omega$ is the volume of the unit cell.}
\end{figure*}
Here we study the influence of the number of MgO layers on the electronic and vibrational properties. As a general trend, increasing the MgO coverage has the effect of isolating the adatom from the interactions with electrons and phonons of the silver substrate. However, the trend is not monotonic due to a marked difference between the geometric configurations with even and odd number of MgO layers. \figurename~\ref{fig:adsorption} shows schematically that when the number of MgO layers is odd, the adatom has an atom of the first silver layer underneath in the same vertical line, whereas it lies on over a hollow site of the Ag(100) surface termination when the number of MgO layers is even. This observation is crucial for understanding the electronic hybridization and the vibrational structure of the system for different coverages of MgO.

\begin{figure}[tbph!]
	\centering
	\includegraphics[width=\linewidth]{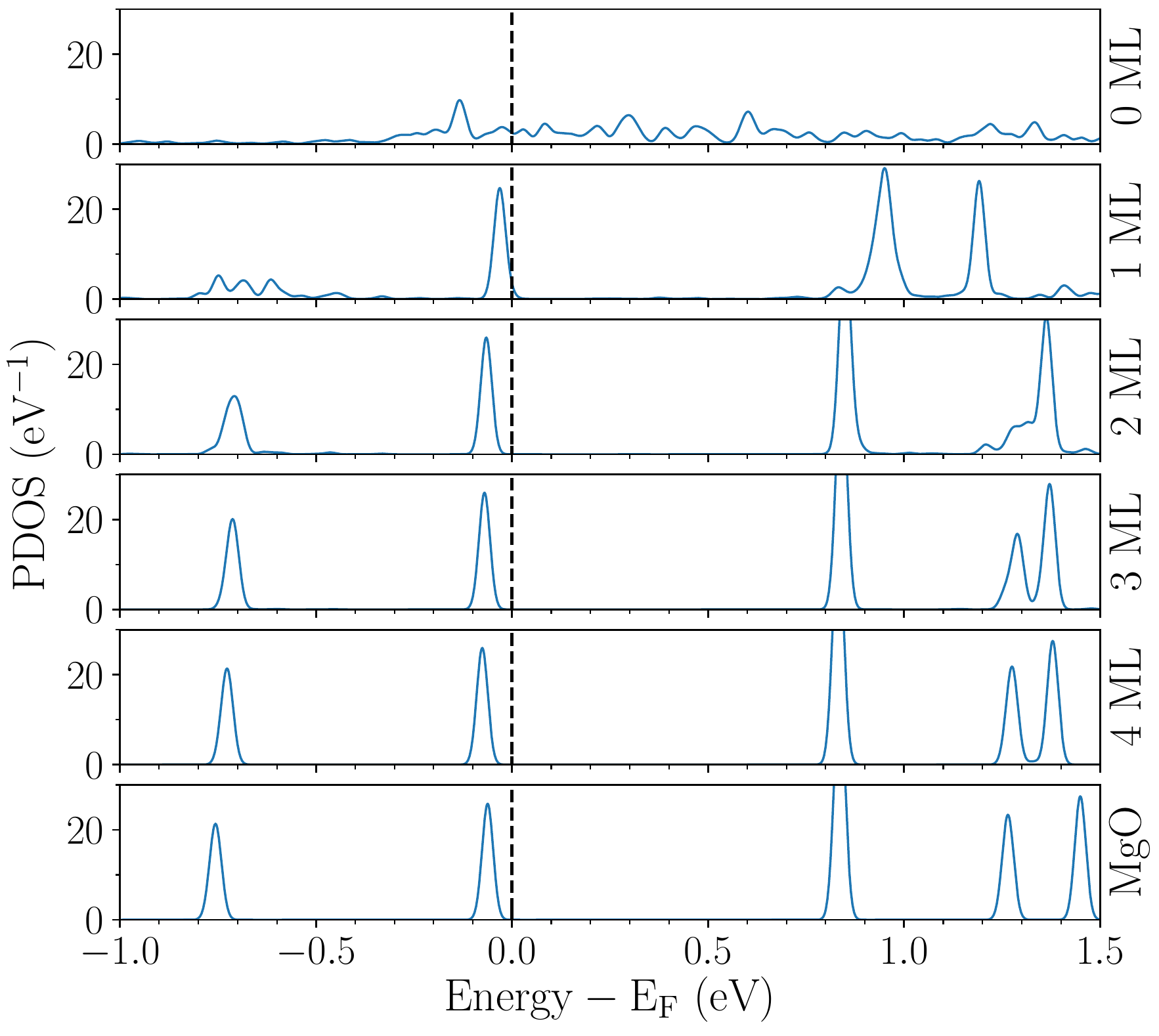}
	\caption{\label{fig:pdosfe_layer} Calculated DOS of Fe-MgO/Ag(100) projected onto the iron adatom for different coverages of MgO and Fe on free standing MgO. The vertical dashed lines indicate the Fermi level.}
\end{figure}

In Figure~\ref{fig:pdosfe_layer} we illustrate the electronic DOS projected on the minority spin channel of the iron adatom for coverages of MgO ranging from 0 to 4~MLs together with iron on free standing MgO. As a general trend we observe that the broadening of the peaks decreases when increasing the number of MgO layers. For the clean Ag(100) surface (0~ML), the iron projected states are completely broadened around the Fermi level, with no clear peak structure. Already for 1~ML coverage, the $3d_{xy}$ peak is localized close to $E_F$, and its width decreases with increasing coverage as a consequence of the insulating nature of MgO. For free standing MgO the iron $3d$ states are located at slightly different energies, but the broadening of the peaks is very similar to the 4~ML MgO/Ag(100) system, indicating that the iron adatom is well protected from the silver substrate electrons with 4~ML of MgO.

An important contribution to the hybridization can be associated to the interface state discussed in \figurename~\ref{fig:bands-slab}, since it lies close to the iron's $3d_{xy}$ state both in real space and in energy. This is made clear in \figurename~\ref{fig:isosurface}, where we show the isosurface of this interface state for different MgO coverages. For 1~ML of MgO, it is clear from \figurename~\ref{fig:isosurface_1ML} that the interface state is mainly localized in the first three layers of silver, but it already shows a considerable hybridization with the iron adatom. For larger coverages of MgO, the hybridization is reduced considerably (note that the value of the isosurface is substantially reduced in order to make the hybridization visible). Another feature revealed by the figure is the different atomic nature of the hybridization depending on the MgO coverage. Due to the qualitative difference between odd and even configurations, the hybridized interface state has a $3d_{z^2}$ orbital character around the iron adatom for an even number of MgO layers, while for odd coverages the hybridization has a $3d_{xz/yz}$ orbital nature on the adatom.

\begin{figure}[tbph!]
	\centering
	\includegraphics[width=\linewidth]{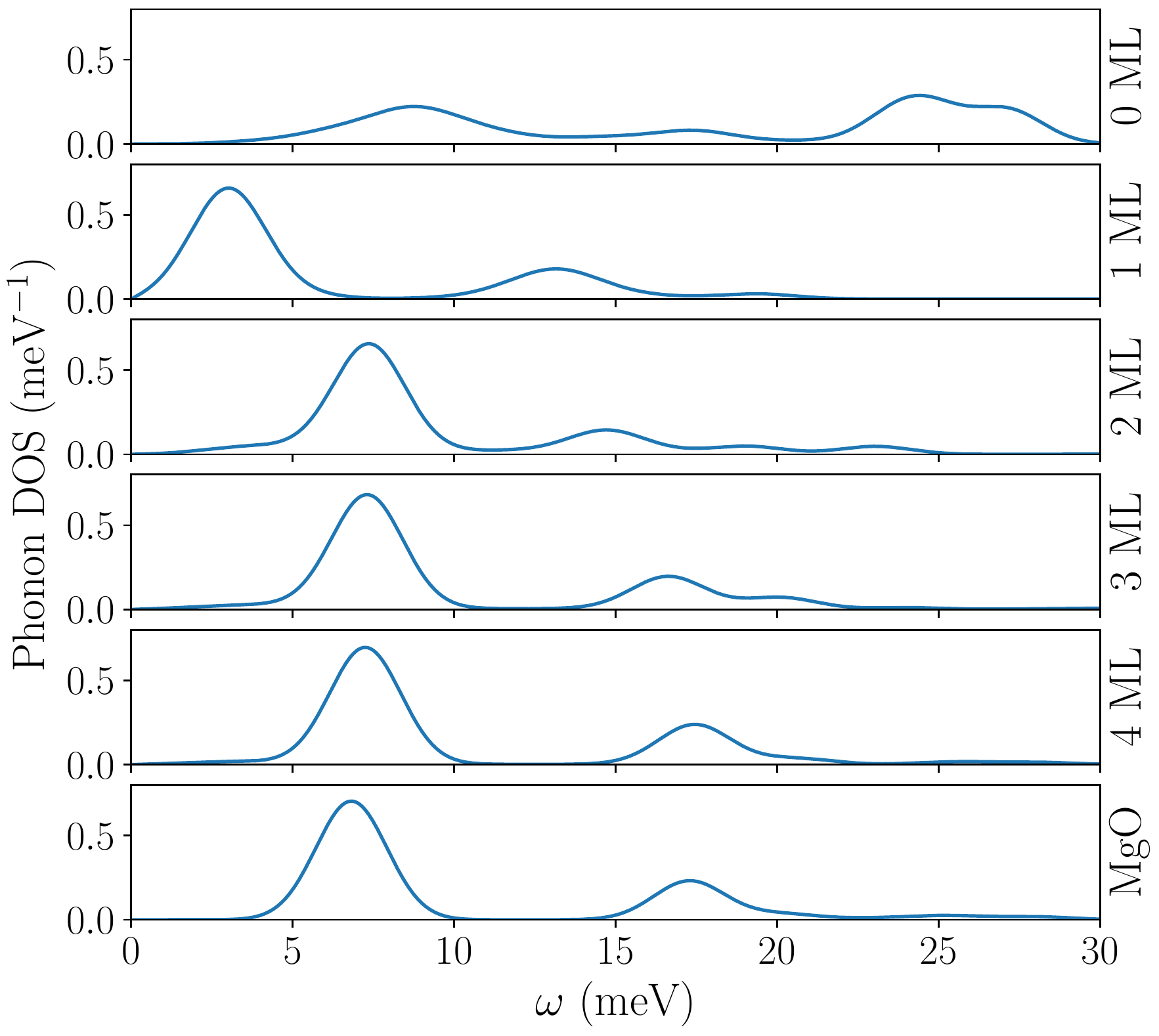}
	\caption{\label{fig:ph_dos_fe_layer} The calculated phonon DOS projected onto the Fe adatom deposited on MgO/Ag(100) with coverages ranging from 0~ML (bare silver surface) to 4~ML of MgO and Fe on free standing MgO.}
\end{figure}
Coming next to the vibrational structure, \figurename~\ref{fig:ph_dos_fe_layer} shows the phonon DOS projected on the adatom for different number of MgO layers. For the system with the iron adatom adsorbed on the clean Ag(100) surface (0~ML), the figure shows that the energy of all vibrational modes is increased compared to coverages $\ge$~1~ML. In the clean substrate the adatom is adsorbed on a hollow site and, thus, it is integrated more compactly into the surface, shifting up the energy of the modes due to the stronger interaction with the neighboring atoms. When adding MgO for coverages $\geq$~2~ML, the largest contribution to the phonon DOS of iron comes from the in-plane modes located approximately at the same energy of around 7.3~meV. In the specific case of a single layer of MgO, the proximity of the silver substrate affects considerably the energy of the localized modes of the iron adatom, softening the energy of the in-plane mode to $\approx$~3.2~meV. On the other hand, our calculations for Fe on free standing MgO show that the in-plane mode energy is approximately at 7~meV, with a phonon DOS practically equal to the 4~ML MgO/Ag(100) system. This  indicates that the vibrational structure will remain unaltered for larger MgO coverages. Incidentally, we note that the energy of the in-plane mode of a Ho adatom in free standing MgO was found at 4.7~meV~\cite{Donati2020}; the ratio between the two energy modes is described reasonably well by the mass ratio of the two adatoms, $\sqrt{M_\text{Fe}/M_\text{Ho}}$. Finally, \figurename~\ref{fig:ph_dos_fe_layer} shows that the out-of plane modes, located between 10 to 20~meV, are hardened by increasing the MgO coverage, the reason being that MgO vibrationenergies are about 4 times more energetic than in silver.

\subsection{Electron-phonon coupling}\label{sec:electron-phonon-coupling}

Having analyzed all the ingredients needed for computing the electron-phonon matrix elements, in this section we study the effect of the electron-phonon interaction on the one-particle states of the system. The size of the electron-phonon interaction is determined by the scattering matrix elements defined in Eqs.~\eqref{eq:matrix-element1} and \eqref{eq:matrix-element2}. Those expressions show that the strength of the electron-phonon interaction depends on the overlap between three quantities: initial and final electronic states, and the potential induced by the phonon. Given that the electron-phonon interaction influences the electronic properties mostly on a energy window on the scale of meV around the Fermi level, the effect on the adatom will be dominated by the $3d_{xy}$ spin-down state. For this reason, and given that the initial $3d_{xy}$ state is localized on the adatom and the only energetically available final states are on the silver substrate, the larger the hybridization of silver states with iron, the larger the electron-phonon scattering matrix elements will be. Furthermore, the vibrational modes that involve the atoms around iron will also be crucial due to the overlap of the potential induced by this phonons with the initial $3d_{xy}$ state.

\subsubsection{Eliashberg function}\label{subsec:eliashberg-function}

The fundamental quantity analyzed in this section is the state-dependent Eliashberg function,
\begin{equation}\label{eq:eliashberg}
	\alpha^2F_i(\omega)
	=
	\sum_{\eta, f} \left| g^{\eta}_{if} \right|^2 \delta(\varepsilon_i-\varepsilon_f) \delta(\omega-\omega_\eta)
.\end{equation}
This quantity is widely used to study the electron-phonon interaction in metals~\cite{Grimvall1981,Mahan2000}, and represents the scattering probability from an initial state with energy $\varepsilon_i$ via a phonon of energy $\omega$. Therefore, this function enables us to identify the phonons that interact most prominently.

\begin{figure}[tbph!]
	\centering
	\includegraphics[width=\linewidth]{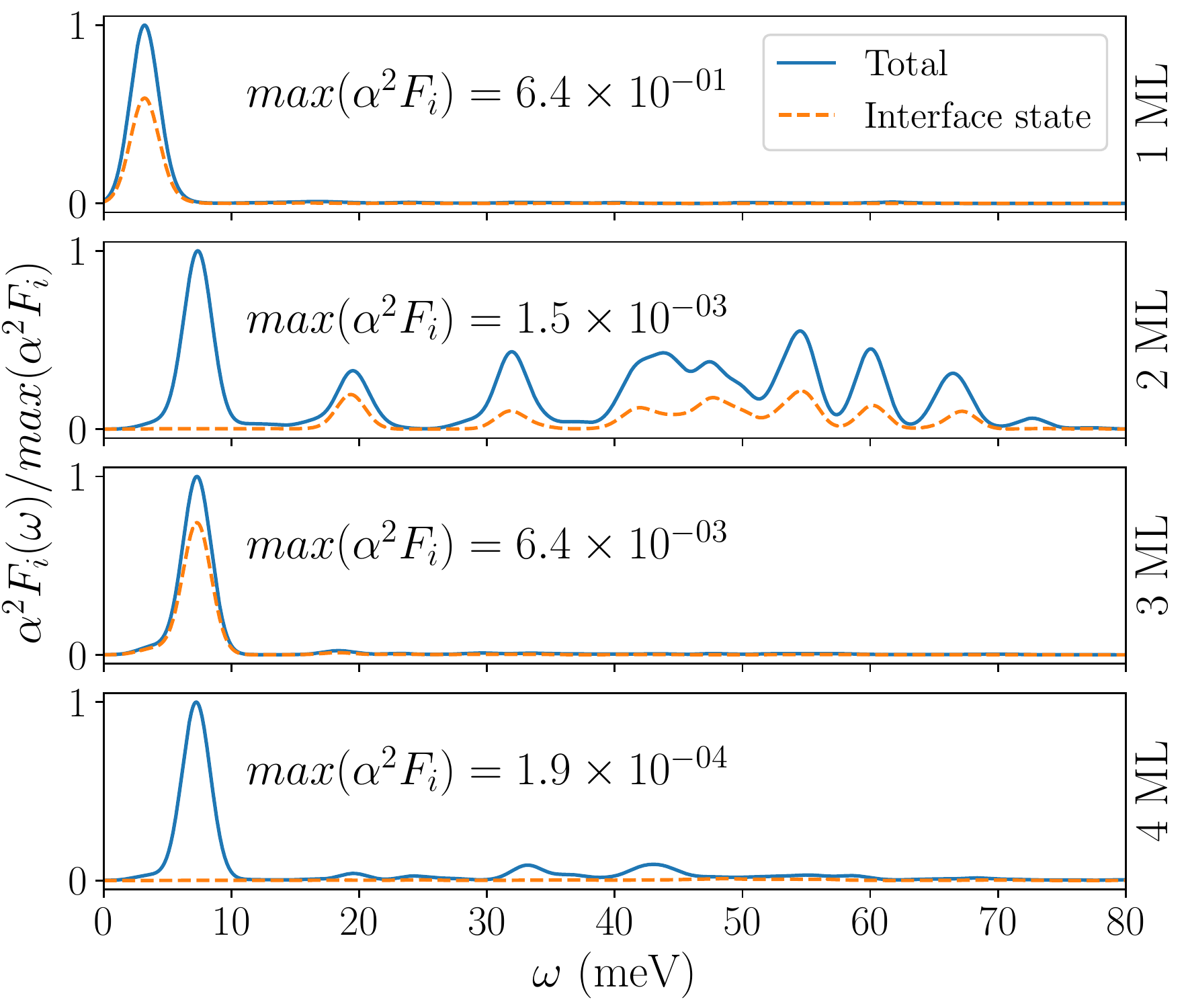}
	\caption{\label{fig:eliashberg_layer} Calculated Eliashberg function of the initial $3d_{xy}$ state of the iron adatom on MgO/Ag(100) for MgO coverages from 1~ML to 4~ML. The solid blue line represents the total scattering probability, while the dashed orange line includes scattering only through the interface state.}
\end{figure}
Figure~\ref{fig:eliashberg_layer} shows our calculated Eliashberg function for the $3d_{xy}$ state of the adatom on the MgO/Ag(100) surface, for MgO coverages ranging from 1~ML to 4~ML (we have not included free standing MgO since the initial $3d_{xy}$ state does not have available final states to scatter to due to the energy gap of MgO around the Fermi level). It is largely dominated by a strong peak located between 3 and 8~meV depending on the coverage, which coincides with the energy of the in-plane phonon modes localized in iron (see \figurename~\ref{fig:ph_dos_fe_layer}). The importance of local in-plane modes revealed by these results is consistent with what was found by~\citet{Donati2020} in a similar system.

Noteworthily, our calculations reveal that for odd coverages of MgO, the main scattering channel is the interface state shown in \figurename~\ref{fig:isosurface}, whose contribution to the Eliashberg function represents as much as the 70\% in the case of 3~ML coverage. For 1~ML, the proximity of the adatom to the silver substrate opens new scattering channels to the bulk silver states, reducing the contribution of the interface state to a 50\% of the total electron-phonon scattering rate. On the other hand, the different atomic nature of the interface state around the adatom for even and odd coverages of MgO (see \figurename~\ref{fig:isosurface}) strongly suppresses scattering to the interface state for an even number of MgO layers, reducing significantly the strong peak observed for an odd coverage of MgO. This makes the contribution of high energy MgO phonons comparable to the contribution of the in-plane modes.

\subsubsection{\texorpdfstring{$\lambda$}{Lambda} parameter and quasiparticle lifetime}\label{subsec:lambda-parameter-and-quasiparticle-lifetime}

The so-called mass enhancement parameter $\lambda$ is a dimensionless parameter which is widely used to characterize the strength of the electron-phonon coupling in metals, and using the Eliashberg function is defined by
\begin{equation}
	\lambda_i
	=
	2\int_0^{\omega_{max}} \frac{\alpha^2F_i(\omega)}{\omega} d\omega
,\end{equation}
for a given state $i$. In metals with full translational symmetry, this quantity describes the mass enhancement of electron quasiparticles at the Fermi level for low temperatures. In the system that we are analyzing, as the most interesting electron states are localized around the iron adatom, we were forced to find another physical interpretation. In fact, it is easily shown that in this case the $\lambda$ parameter describes the energy shift due to the electron-phonon interaction (see \appendixname~\ref{sec:energy-renormalization-of-localized-electronic-states}),
\begin{align}\label{eq:energy-shift}
	\Delta\varepsilon
	\approx
	-\varepsilon_0\lambda \left( \frac{\omega_0}{\varepsilon_0} \right)^2
	\text{for $|\varepsilon_0|\gg\omega_0$}
,\end{align}
and provides a feeling of the strength of the electron-phonon coupling. Above, $\varepsilon_0$ and $\omega_0$ represent the unperturbed  energy of the localized state and the energy of the localized vibrational mode, respectively.

\begin{figure}[tbph!]
	\centering
	\includegraphics[width=\linewidth]{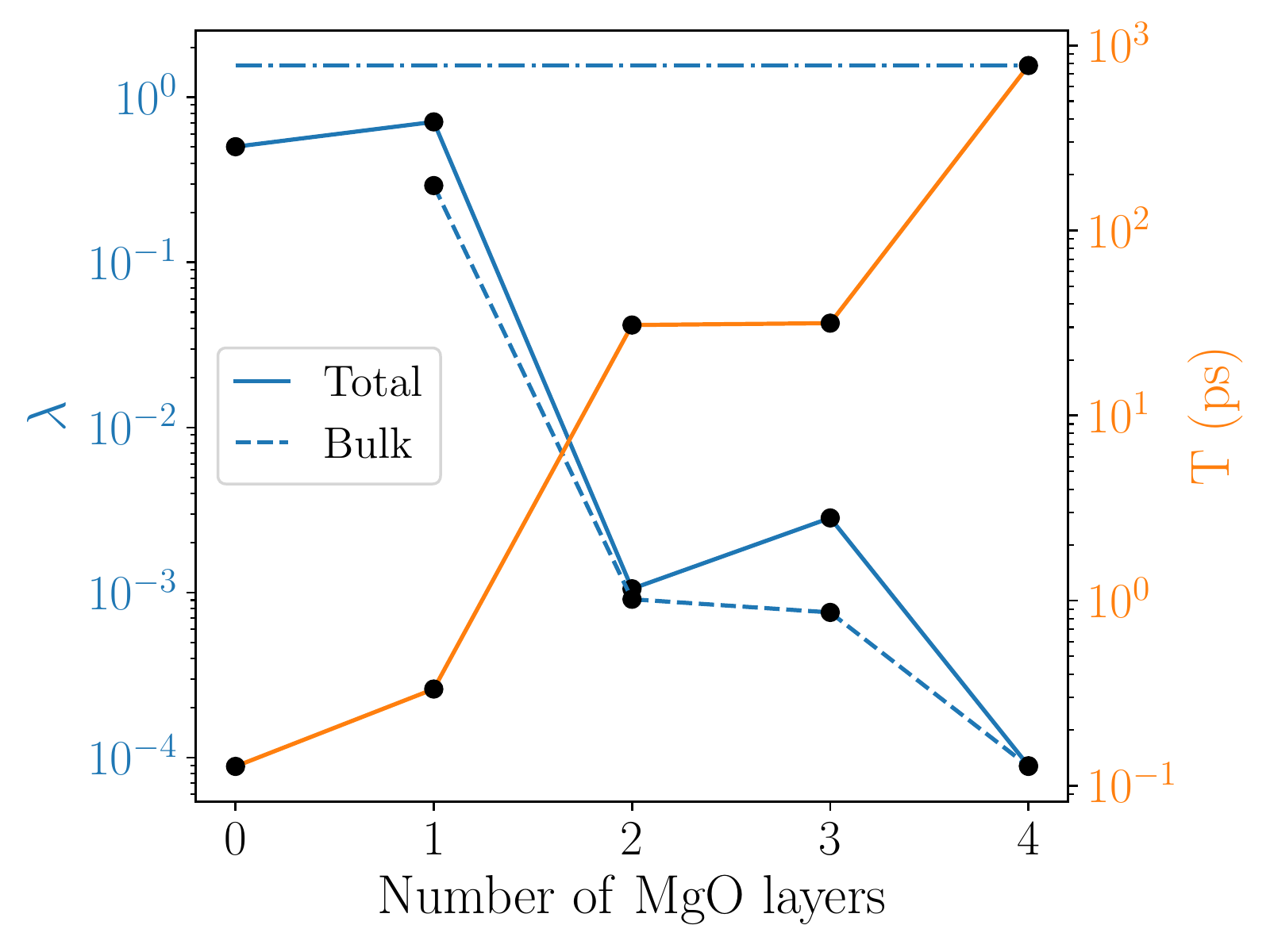}
	\caption{\label{fig:lifetime_lambda} Calculated electron-phonon $\lambda$ parameter (left axis) and electron-phonon lifetime of a one-particle excitation (right axis) for the $3d_{xy}$ orbital of the iron adatom deposited on MgO/Ag(100) as a function of the MgO coverage. The dashed line represents the $\lambda$ parameter for scattering through silver bulk states, whereas the solid line represents the total $\lambda$ parameter (silver bulk plus interface state scattering channels). The dash-dotted line indicates the $\lambda$ parameter of bulk Pb, a material with strong electron-phonon coupling.}
\end{figure}

The calculated $\lambda$ parameter of the localized iron $3d_{xy}$ state is shown in \figurename~\ref{fig:lifetime_lambda} for the different MgO coverages; in order to have a comparative indication of the strength of the coupling we have also included the $\lambda$ parameter of bulk Pb, which ranks among the strongest ever reported with $\lambda=1.56$~\cite{Sklyadneva2012}, whereas on noble metal surfaces it can vary on the range \mbox{$0.11-0.01$}~\cite{Hofmann2009}. We observe a clear step between 1~ML and 2~ML coverages. In fact, for $\leq$~1~ML coverages, the strength of the electron-phonon interaction is quantitatively close to the one found for bulk Pb, which from Eq.~\eqref{eq:energy-shift} gives a positive energy shift of around 10\% for the $3d_{xy}$ state. We find that for larger coverages, the $\lambda$ parameter shows a reduction of two orders of magnitude, indicating that the $3d_{xy}$ orbital becomes effectively protected from the electron-phonon interaction. Clearly, our analysis indicates that a single MgO layer is not enough to screen or decouple the electronic states of the iron adatom from the silver substrate. Also, we observe that the layer dependence of the $\lambda$ parameter follows a clear odd-even step structure, whereas it decreases if we do not consider scattering through the interface state (see dashed line in \figurename~\ref{fig:lifetime_lambda}). Thus, the interface state is the main responsible of the step structure, which is consistent with our analysis of the geometric configuration in terms of the number of MgO layers and the connection with the hybridization of the interface state.

The Eliashberg function also allows us to calculate the lifetime of excited single electron quasiparticles, integrating the scattering probabilities at all phonon energies,
\begin{equation}
	\tau_i^{-1} = \Gamma_i = 2\pi\int_0^{\omega_{max}} \alpha^2F_i(\omega) d\omega
,\end{equation}
which resembles Fermi's Golden Rule. Figure~\ref{fig:lifetime_lambda} shows our calculated lifetime for a hole on the $3d_{xy}$ orbital of the adatom as a function of MgO layers. For 0~ML and 1~ML coverages, the lifetime of the quasiparticle is of the order of 0.1~ps, whereas we obtain a lifetime of around 30 ps for 2 and 3~ML coverages and of 800~ps for 4~MLs. Again, we observe a clear step structure, with a jump of two orders of magnitude from 1~ML to 2~ML coverage, which is also caused by the absence of the interface state scattering channel for an even number of MgO layers.

\section{Summary and outlook}\label{sec:conclusions}

We have conducted a detailed \textit{ab-initio} analysis of the spin-diagonal electron-phonon interaction of an iron adatom on the MgO/Ag(100) surface. We have calculated the electronic and vibrational structures, pointing out the importance of the silver substrate providing final states to scatter on the electron-phonon interaction. We have computed the strength of the electron-phonon coupling, demonstrating by considering the Eliashberg function that the in-plane oscillations of the adatom are the most important vibrational modes, in good agreement with what was found in Ho on MgO/Ag(100)~\cite{Donati2020}. Moreover, we have found that the strength of the electron-phonon interaction shows qualitative differences for even and odd number of MgO layers, as deduced by the calculated $\lambda$ parameter and quasiparticle lifetime. This difference between even and odd coverages is explained by the presence of an interface state of the substrate close to the high symmetry point $X$, which represents the most important scattering channel for odd coverages of MgO; its contribution represents up to 70\% of the total scattering rate in the case of 3~ML coverage, whereas for an even number of MgO layers its contribution is highly suppressed.

As a central result of our work, we have shown that a single MgO layer is not capable of effectively screening the spin-diagonal electron-phonon interaction on the iron adatom, whose calculated strength is comparable to the largest values found among bulk materials. In turn, this scattering channel is deeply suppressed for two or more layers of MgO. Interestingly, a recent spin-polarized scanning tunneling microscope experiment on the same system reported a long-living magnetic moment of iron \emph{only} for $\geq$~2~ML MgO coverages~\cite{Paul2017}, and the lifetime of the magnetic moment has never been reported for smaller coverages. While a single monolayer of MgO on Ag(100) occurs infrequently in experiment~\cite{Paul2017}, this coincidence might also suggests that an effective screening of the electron-phonon scattering channel studied in this work is an important ingredient for achieving magnetic stability of single adatoms.

Beyond this contribution, scattering terms that flip the spin of the adatom electron states are required in order to gain further insight into the phonon-driven relaxation mechanism. The spin-flip electron-phonon contribution is mediated by the spin-orbit interaction and determines the phonon spin relaxation time (not to be confused with the quasiparticle lifetime reported in the present work). In molecular magnets, this quantity has been calculated by combining \textit{ab-initio} calculations with spin-phonon dynamics~\cite{Lunghi2017},	but no similar study has been reported in single adatoms to our knowledge. While in this work we have shown that the spin-diagonal contribution depends strongly on the number of MgO layers, the spin-flip process is expected to be a more local effect as it takes place in the adatom itself, which is likely to be less influenced by the MgO coverage. These and further aspects of the spin-flip electron-phonon interaction will be the subject of a future work.

\section*{Acknowledgments}

The authors acknowledge the Department of Education, Universities and Research of the Eusko Jaurlaritza and the University of the Basque Country UPV/EHU (Grant No. IT1260-19), the Spanish Ministry of Economy and Competitiveness MINECO (Grants No. FIS2016-75862-P and No. PID2019-103910GB-I00), and the University of the Basque Country UPV/EHU (Grant No. GIU18/138) for financial support. This project has also received funding from the  European Union's Horizon 2020 research and innovation programme under the Marie Sklodowska-Curie grant agreement No. 839237 and the European Research Council (ERC) grant agreement No. 946629. H.G.-M. acknowledges the Spanish Ministry of Economy and Competitiveness MINECO (Grant No. BES-2017-080039) and the Donostia International Physics Center (DIPC) for financial support. Computer facilities were provided by the DIPC and Centro de F{\'i}sica de Materiales.

\appendix

\section{Energy renormalization of localized electronic states}\label{sec:energy-renormalization-of-localized-electronic-states}

In this Appendix we will derive the energy shift that experiences a localized electronic state in the Einstein model. In this case, the energy renormalization of the electronic state can be directly related with the $\lambda$ parameter. In the Einsten model, with a dispersionless phonon with energy $\omega_0$, the real part of the electron-phonon self-energy of an electron is determined by the $\lambda$ parameter:
\begin{equation}
	\Sigma(\omega)
	=
	\lambda\frac{\omega_0}{2} log\left| \frac{\omega-\omega_0}{\omega+\omega_0} \right|
.\end{equation}
The relevant vibrational modes in Fe on MgO/Ag(100) are the localized dispersionless in-plane modes of the iron adatom, which are very soft modes. Therefore, the energy of the localized electronic $3d_{xy}$ state is much higher, and the self-energy can be safely approximated as
\begin{equation}
	\Sigma(\omega)
	\approx
	-\lambda \frac{\omega_0^2}{\omega} \text{for $\omega \gg \omega_0$}
.\end{equation}
This makes possible to solve the Dyson equation analytically for the $3d_{xy}$ orbital,
\begin{equation}
	\varepsilon
	=
	\varepsilon_0 + \Sigma(\varepsilon)
	=
	\varepsilon_0 - \lambda\frac{\omega_0^2}{\varepsilon}
,\end{equation}
and obtain the renormalized energy $\varepsilon$ in terms of the phonon energy $\omega_0$, the $\lambda$ parameter and the unperturbed electron energy $\varepsilon_0$:
\begin{align}
	\varepsilon
	&=
	\frac{\varepsilon_0}{2}
	\left[
	1 + \sqrt{1-4\lambda \left( \frac{\omega_0}{\varepsilon_0} \right)^2}
	\right]\nonumber\\
	&\approx
	\varepsilon_0
	\left[ 1 - \lambda \left( \frac{\omega_0}{\varepsilon_0} \right)^2 \right]
	\text{for $|\varepsilon_0|\gg\omega_0$}
.\end{align}
Thus, the shift in energy is
\begin{align}
	\Delta\varepsilon
	\approx
	-\varepsilon_0\lambda \left( \frac{\omega_0}{\varepsilon_0} \right)^2
	\text{for $|\varepsilon_0|\gg\omega_0$}
,\end{align}
which always moves the energy of the localized state closer to the Fermi level. This last formula is shown in the main text in Eq.~\eqref{eq:energy-shift}.

\bibliographystyle{apsrev4-2}
\bibliography{./bibliografia}

\begin{thebibliography}{62}%
\makeatletter
\providecommand \@ifxundefined [1]{%
 \@ifx{#1\undefined}
}%
\providecommand \@ifnum [1]{%
 \ifnum #1\expandafter \@firstoftwo
 \else \expandafter \@secondoftwo
 \fi
}%
\providecommand \@ifx [1]{%
 \ifx #1\expandafter \@firstoftwo
 \else \expandafter \@secondoftwo
 \fi
}%
\providecommand \natexlab [1]{#1}%
\providecommand \enquote  [1]{``#1''}%
\providecommand \bibnamefont  [1]{#1}%
\providecommand \bibfnamefont [1]{#1}%
\providecommand \citenamefont [1]{#1}%
\providecommand \href@noop [0]{\@secondoftwo}%
\providecommand \href [0]{\begingroup \@sanitize@url \@href}%
\providecommand \@href[1]{\@@startlink{#1}\@@href}%
\providecommand \@@href[1]{\endgroup#1\@@endlink}%
\providecommand \@sanitize@url [0]{\catcode `\\12\catcode `\$12\catcode
  `\&12\catcode `\#12\catcode `\^12\catcode `\_12\catcode `\%12\relax}%
\providecommand \@@startlink[1]{}%
\providecommand \@@endlink[0]{}%
\providecommand \url  [0]{\begingroup\@sanitize@url \@url }%
\providecommand \@url [1]{\endgroup\@href {#1}{\urlprefix }}%
\providecommand \urlprefix  [0]{URL }%
\providecommand \Eprint [0]{\href }%
\providecommand \doibase [0]{https://doi.org/}%
\providecommand \selectlanguage [0]{\@gobble}%
\providecommand \bibinfo  [0]{\@secondoftwo}%
\providecommand \bibfield  [0]{\@secondoftwo}%
\providecommand \translation [1]{[#1]}%
\providecommand \BibitemOpen [0]{}%
\providecommand \bibitemStop [0]{}%
\providecommand \bibitemNoStop [0]{.\EOS\space}%
\providecommand \EOS [0]{\spacefactor3000\relax}%
\providecommand \BibitemShut  [1]{\csname bibitem#1\endcsname}%
\let\auto@bib@innerbib\@empty
\bibitem [{\citenamefont {Natterer}\ \emph {et~al.}(2017)\citenamefont
  {Natterer}, \citenamefont {Yang}, \citenamefont {Paul}, \citenamefont
  {Willke}, \citenamefont {Choi}, \citenamefont {Greber}, \citenamefont
  {Heinrich},\ and\ \citenamefont {Lutz}}]{Natterer2017}%
  \BibitemOpen
  \bibfield  {author} {\bibinfo {author} {\bibfnamefont {F.~D.}\ \bibnamefont
  {Natterer}}, \bibinfo {author} {\bibfnamefont {K.}~\bibnamefont {Yang}},
  \bibinfo {author} {\bibfnamefont {W.}~\bibnamefont {Paul}}, \bibinfo {author}
  {\bibfnamefont {P.}~\bibnamefont {Willke}}, \bibinfo {author} {\bibfnamefont
  {T.}~\bibnamefont {Choi}}, \bibinfo {author} {\bibfnamefont {T.}~\bibnamefont
  {Greber}}, \bibinfo {author} {\bibfnamefont {A.~J.}\ \bibnamefont
  {Heinrich}},\ and\ \bibinfo {author} {\bibfnamefont {C.~P.}\ \bibnamefont
  {Lutz}},\ }\href {https://doi.org/10.1038/nature21371} {\bibfield  {journal}
  {\bibinfo  {journal} {Nature}\ }\textbf {\bibinfo {volume} {543}},\ \bibinfo
  {pages} {226} (\bibinfo {year} {2017})}\BibitemShut {NoStop}%
\bibitem [{\citenamefont {Khajetoorians}\ \emph
  {et~al.}(2011{\natexlab{a}})\citenamefont {Khajetoorians}, \citenamefont
  {Wiebe}, \citenamefont {Chilian},\ and\ \citenamefont
  {Wiesendanger}}]{Khajetoorians2011a}%
  \BibitemOpen
  \bibfield  {author} {\bibinfo {author} {\bibfnamefont {A.~A.}\ \bibnamefont
  {Khajetoorians}}, \bibinfo {author} {\bibfnamefont {J.}~\bibnamefont
  {Wiebe}}, \bibinfo {author} {\bibfnamefont {B.}~\bibnamefont {Chilian}},\
  and\ \bibinfo {author} {\bibfnamefont {R.}~\bibnamefont {Wiesendanger}},\
  }\href {https://doi.org/10.1126/science.1201725} {\bibfield  {journal}
  {\bibinfo  {journal} {Science}\ }\textbf {\bibinfo {volume} {332}},\ \bibinfo
  {pages} {1062} (\bibinfo {year} {2011}{\natexlab{a}})}\BibitemShut {NoStop}%
\bibitem [{\citenamefont {Lang}\ \emph {et~al.}(1994)\citenamefont {Lang},
  \citenamefont {Stepanyuk}, \citenamefont {Wildberger}, \citenamefont
  {Zeller},\ and\ \citenamefont {Dederichs}}]{Lang1994}%
  \BibitemOpen
  \bibfield  {author} {\bibinfo {author} {\bibfnamefont {P.}~\bibnamefont
  {Lang}}, \bibinfo {author} {\bibfnamefont {V.~S.}\ \bibnamefont {Stepanyuk}},
  \bibinfo {author} {\bibfnamefont {K.}~\bibnamefont {Wildberger}}, \bibinfo
  {author} {\bibfnamefont {R.}~\bibnamefont {Zeller}},\ and\ \bibinfo {author}
  {\bibfnamefont {P.~H.}\ \bibnamefont {Dederichs}},\ }\href
  {https://doi.org/10.1016/0038-1098(94)90767-6} {\bibfield  {journal}
  {\bibinfo  {journal} {Solid State Communications}\ }\textbf {\bibinfo
  {volume} {92}},\ \bibinfo {pages} {755} (\bibinfo {year} {1994})}\BibitemShut
  {NoStop}%
\bibitem [{\citenamefont {Lounis}\ \emph {et~al.}(2010)\citenamefont {Lounis},
  \citenamefont {Costa}, \citenamefont {Muniz},\ and\ \citenamefont
  {Mills}}]{Lounis2010}%
  \BibitemOpen
  \bibfield  {author} {\bibinfo {author} {\bibfnamefont {S.}~\bibnamefont
  {Lounis}}, \bibinfo {author} {\bibfnamefont {A.~T.}\ \bibnamefont {Costa}},
  \bibinfo {author} {\bibfnamefont {R.~B.}\ \bibnamefont {Muniz}},\ and\
  \bibinfo {author} {\bibfnamefont {D.~L.}\ \bibnamefont {Mills}},\ }\href
  {https://doi.org/10.1103/PhysRevLett.105.187205} {\bibfield  {journal}
  {\bibinfo  {journal} {Physical Review Letters}\ }\textbf {\bibinfo {volume}
  {105}},\ \bibinfo {pages} {187205} (\bibinfo {year} {2010})}\BibitemShut
  {NoStop}%
\bibitem [{\citenamefont {Iba{\~{n}}ez-Azpiroz}\ \emph
  {et~al.}(2017{\natexlab{a}})\citenamefont {Iba{\~{n}}ez-Azpiroz},
  \citenamefont {{dos Santos Dias}}, \citenamefont {Bl{\"{u}}gel},\ and\
  \citenamefont {Lounis}}]{Ibanez-Azpiroz2017a}%
  \BibitemOpen
  \bibfield  {author} {\bibinfo {author} {\bibfnamefont {J.}~\bibnamefont
  {Iba{\~{n}}ez-Azpiroz}}, \bibinfo {author} {\bibfnamefont {M.}~\bibnamefont
  {{dos Santos Dias}}}, \bibinfo {author} {\bibfnamefont {S.}~\bibnamefont
  {Bl{\"{u}}gel}},\ and\ \bibinfo {author} {\bibfnamefont {S.}~\bibnamefont
  {Lounis}},\ }\href {https://doi.org/10.1103/PhysRevB.96.144410} {\bibfield
  {journal} {\bibinfo  {journal} {Physical Review B}\ }\textbf {\bibinfo
  {volume} {96}},\ \bibinfo {pages} {144410} (\bibinfo {year}
  {2017}{\natexlab{a}})}\BibitemShut {NoStop}%
\bibitem [{\citenamefont {Lorente}\ and\ \citenamefont
  {Gauyacq}(2009)}]{Lorente2009}%
  \BibitemOpen
  \bibfield  {author} {\bibinfo {author} {\bibfnamefont {N.}~\bibnamefont
  {Lorente}}\ and\ \bibinfo {author} {\bibfnamefont {J.-P.}\ \bibnamefont
  {Gauyacq}},\ }\href {https://doi.org/10.1103/PhysRevLett.103.176601}
  {\bibfield  {journal} {\bibinfo  {journal} {Physical Review Letters}\
  }\textbf {\bibinfo {volume} {103}},\ \bibinfo {pages} {176601} (\bibinfo
  {year} {2009})}\BibitemShut {NoStop}%
\bibitem [{\citenamefont {Delgado}\ and\ \citenamefont
  {Fern{\'{a}}ndez-Rossier}(2017)}]{Delgado2017}%
  \BibitemOpen
  \bibfield  {author} {\bibinfo {author} {\bibfnamefont {F.}~\bibnamefont
  {Delgado}}\ and\ \bibinfo {author} {\bibfnamefont {J.}~\bibnamefont
  {Fern{\'{a}}ndez-Rossier}},\ }\href
  {https://doi.org/10.1016/j.progsurf.2016.12.001} {\bibfield  {journal}
  {\bibinfo  {journal} {Progress in Surface Science}\ }\textbf {\bibinfo
  {volume} {92}},\ \bibinfo {pages} {40} (\bibinfo {year} {2017})}\BibitemShut
  {NoStop}%
\bibitem [{\citenamefont {Wolf}\ \emph {et~al.}(2020)\citenamefont {Wolf},
  \citenamefont {Delgado}, \citenamefont {Reina},\ and\ \citenamefont
  {Lorente}}]{Wolf2020}%
  \BibitemOpen
  \bibfield  {author} {\bibinfo {author} {\bibfnamefont {C.}~\bibnamefont
  {Wolf}}, \bibinfo {author} {\bibfnamefont {F.}~\bibnamefont {Delgado}},
  \bibinfo {author} {\bibfnamefont {J.}~\bibnamefont {Reina}},\ and\ \bibinfo
  {author} {\bibfnamefont {N.}~\bibnamefont {Lorente}},\ }\href
  {https://doi.org/10.1021/acs.jpca.9b10749} {\bibfield  {journal} {\bibinfo
  {journal} {The Journal of Physical Chemistry A}\ }\textbf {\bibinfo {volume}
  {124}},\ \bibinfo {pages} {2318} (\bibinfo {year} {2020})}\BibitemShut
  {NoStop}%
\bibitem [{\citenamefont {Heinrich}\ \emph {et~al.}(2004)\citenamefont
  {Heinrich}, \citenamefont {Gupta}, \citenamefont {Lutz},\ and\ \citenamefont
  {Eigler}}]{Heinrich2004}%
  \BibitemOpen
  \bibfield  {author} {\bibinfo {author} {\bibfnamefont {A.~J.}\ \bibnamefont
  {Heinrich}}, \bibinfo {author} {\bibfnamefont {J.~A.}\ \bibnamefont {Gupta}},
  \bibinfo {author} {\bibfnamefont {C.~P.}\ \bibnamefont {Lutz}},\ and\
  \bibinfo {author} {\bibfnamefont {D.~M.}\ \bibnamefont {Eigler}},\ }\href
  {https://doi.org/10.1126/science.1101077} {\bibfield  {journal} {\bibinfo
  {journal} {Science}\ }\textbf {\bibinfo {volume} {306}},\ \bibinfo {pages}
  {466} (\bibinfo {year} {2004})}\BibitemShut {NoStop}%
\bibitem [{\citenamefont {Hirjibehedin}\ \emph {et~al.}(2006)\citenamefont
  {Hirjibehedin}, \citenamefont {Lutz},\ and\ \citenamefont
  {Heinrich}}]{Hirjibehedin2006}%
  \BibitemOpen
  \bibfield  {author} {\bibinfo {author} {\bibfnamefont {C.~F.}\ \bibnamefont
  {Hirjibehedin}}, \bibinfo {author} {\bibfnamefont {C.~P.}\ \bibnamefont
  {Lutz}},\ and\ \bibinfo {author} {\bibfnamefont {A.~J.}\ \bibnamefont
  {Heinrich}},\ }\href {https://doi.org/10.1126/science.1125398} {\bibfield
  {journal} {\bibinfo  {journal} {Science}\ }\textbf {\bibinfo {volume}
  {312}},\ \bibinfo {pages} {1021} (\bibinfo {year} {2006})}\BibitemShut
  {NoStop}%
\bibitem [{\citenamefont {Hirjibehedin}\ \emph {et~al.}(2007)\citenamefont
  {Hirjibehedin}, \citenamefont {Lin}, \citenamefont {Otte}, \citenamefont
  {Ternes}, \citenamefont {Lutz}, \citenamefont {Jones},\ and\ \citenamefont
  {Heinrich}}]{Hirjibehedin2007}%
  \BibitemOpen
  \bibfield  {author} {\bibinfo {author} {\bibfnamefont {C.~F.}\ \bibnamefont
  {Hirjibehedin}}, \bibinfo {author} {\bibfnamefont {C.-Y.}\ \bibnamefont
  {Lin}}, \bibinfo {author} {\bibfnamefont {A.~F.}\ \bibnamefont {Otte}},
  \bibinfo {author} {\bibfnamefont {M.}~\bibnamefont {Ternes}}, \bibinfo
  {author} {\bibfnamefont {C.~P.}\ \bibnamefont {Lutz}}, \bibinfo {author}
  {\bibfnamefont {B.~A.}\ \bibnamefont {Jones}},\ and\ \bibinfo {author}
  {\bibfnamefont {A.~J.}\ \bibnamefont {Heinrich}},\ }\href
  {https://doi.org/10.1126/science.1146110} {\bibfield  {journal} {\bibinfo
  {journal} {Science}\ }\textbf {\bibinfo {volume} {317}},\ \bibinfo {pages}
  {1199} (\bibinfo {year} {2007})}\BibitemShut {NoStop}%
\bibitem [{\citenamefont {Khajetoorians}\ \emph
  {et~al.}(2011{\natexlab{b}})\citenamefont {Khajetoorians}, \citenamefont
  {Lounis}, \citenamefont {Chilian}, \citenamefont {Costa}, \citenamefont
  {Zhou}, \citenamefont {Mills}, \citenamefont {Wiebe},\ and\ \citenamefont
  {Wiesendanger}}]{Khajetoorians2011}%
  \BibitemOpen
  \bibfield  {author} {\bibinfo {author} {\bibfnamefont {A.~A.}\ \bibnamefont
  {Khajetoorians}}, \bibinfo {author} {\bibfnamefont {S.}~\bibnamefont
  {Lounis}}, \bibinfo {author} {\bibfnamefont {B.}~\bibnamefont {Chilian}},
  \bibinfo {author} {\bibfnamefont {A.~T.}\ \bibnamefont {Costa}}, \bibinfo
  {author} {\bibfnamefont {L.}~\bibnamefont {Zhou}}, \bibinfo {author}
  {\bibfnamefont {D.~L.}\ \bibnamefont {Mills}}, \bibinfo {author}
  {\bibfnamefont {J.}~\bibnamefont {Wiebe}},\ and\ \bibinfo {author}
  {\bibfnamefont {R.}~\bibnamefont {Wiesendanger}},\ }\href
  {https://doi.org/10.1103/PhysRevLett.106.037205} {\bibfield  {journal}
  {\bibinfo  {journal} {Physical Review Letters}\ }\textbf {\bibinfo {volume}
  {106}},\ \bibinfo {pages} {037205} (\bibinfo {year}
  {2011}{\natexlab{b}})}\BibitemShut {NoStop}%
\bibitem [{\citenamefont {Donati}\ \emph {et~al.}(2013)\citenamefont {Donati},
  \citenamefont {Dubout}, \citenamefont {Aut{\`{e}}s}, \citenamefont {Patthey},
  \citenamefont {Calleja}, \citenamefont {Gambardella}, \citenamefont
  {Yazyev},\ and\ \citenamefont {Brune}}]{Donati2013}%
  \BibitemOpen
  \bibfield  {author} {\bibinfo {author} {\bibfnamefont {F.}~\bibnamefont
  {Donati}}, \bibinfo {author} {\bibfnamefont {Q.}~\bibnamefont {Dubout}},
  \bibinfo {author} {\bibfnamefont {G.}~\bibnamefont {Aut{\`{e}}s}}, \bibinfo
  {author} {\bibfnamefont {F.}~\bibnamefont {Patthey}}, \bibinfo {author}
  {\bibfnamefont {F.}~\bibnamefont {Calleja}}, \bibinfo {author} {\bibfnamefont
  {P.}~\bibnamefont {Gambardella}}, \bibinfo {author} {\bibfnamefont {O.~V.}\
  \bibnamefont {Yazyev}},\ and\ \bibinfo {author} {\bibfnamefont
  {H.}~\bibnamefont {Brune}},\ }\href
  {https://doi.org/10.1103/PhysRevLett.111.236801} {\bibfield  {journal}
  {\bibinfo  {journal} {Physical Review Letters}\ }\textbf {\bibinfo {volume}
  {111}},\ \bibinfo {pages} {236801} (\bibinfo {year} {2013})}\BibitemShut
  {NoStop}%
\bibitem [{\citenamefont {Hermenau}\ \emph {et~al.}(2018)\citenamefont
  {Hermenau}, \citenamefont {Ternes}, \citenamefont {Steinbrecher},
  \citenamefont {Wiesendanger},\ and\ \citenamefont {Wiebe}}]{Hermenau2018}%
  \BibitemOpen
  \bibfield  {author} {\bibinfo {author} {\bibfnamefont {J.}~\bibnamefont
  {Hermenau}}, \bibinfo {author} {\bibfnamefont {M.}~\bibnamefont {Ternes}},
  \bibinfo {author} {\bibfnamefont {M.}~\bibnamefont {Steinbrecher}}, \bibinfo
  {author} {\bibfnamefont {R.}~\bibnamefont {Wiesendanger}},\ and\ \bibinfo
  {author} {\bibfnamefont {J.}~\bibnamefont {Wiebe}},\ }\href
  {https://doi.org/10.1021/acs.nanolett.7b05392} {\bibfield  {journal}
  {\bibinfo  {journal} {Nano Letters}\ }\textbf {\bibinfo {volume} {18}},\
  \bibinfo {pages} {1978} (\bibinfo {year} {2018})}\BibitemShut {NoStop}%
\bibitem [{\citenamefont {Donati}\ \emph
  {et~al.}(2014{\natexlab{a}})\citenamefont {Donati}, \citenamefont
  {Gragnaniello}, \citenamefont {Cavallin}, \citenamefont {Natterer},
  \citenamefont {Dubout}, \citenamefont {Pivetta}, \citenamefont {Patthey},
  \citenamefont {Dreiser}, \citenamefont {Piamonteze}, \citenamefont
  {Rusponi},\ and\ \citenamefont {Brune}}]{Donati2014}%
  \BibitemOpen
  \bibfield  {author} {\bibinfo {author} {\bibfnamefont {F.}~\bibnamefont
  {Donati}}, \bibinfo {author} {\bibfnamefont {L.}~\bibnamefont
  {Gragnaniello}}, \bibinfo {author} {\bibfnamefont {A.}~\bibnamefont
  {Cavallin}}, \bibinfo {author} {\bibfnamefont {F.~D.}\ \bibnamefont
  {Natterer}}, \bibinfo {author} {\bibfnamefont {Q.}~\bibnamefont {Dubout}},
  \bibinfo {author} {\bibfnamefont {M.}~\bibnamefont {Pivetta}}, \bibinfo
  {author} {\bibfnamefont {F.}~\bibnamefont {Patthey}}, \bibinfo {author}
  {\bibfnamefont {J.}~\bibnamefont {Dreiser}}, \bibinfo {author} {\bibfnamefont
  {C.}~\bibnamefont {Piamonteze}}, \bibinfo {author} {\bibfnamefont
  {S.}~\bibnamefont {Rusponi}},\ and\ \bibinfo {author} {\bibfnamefont
  {H.}~\bibnamefont {Brune}},\ }\href
  {https://doi.org/10.1103/PhysRevLett.113.177201} {\bibfield  {journal}
  {\bibinfo  {journal} {Physical Review Letters}\ }\textbf {\bibinfo {volume}
  {113}},\ \bibinfo {pages} {177201} (\bibinfo {year}
  {2014}{\natexlab{a}})}\BibitemShut {NoStop}%
\bibitem [{\citenamefont {Donati}\ \emph
  {et~al.}(2014{\natexlab{b}})\citenamefont {Donati}, \citenamefont {Singha},
  \citenamefont {Stepanow}, \citenamefont {W{\"{a}}ckerlin}, \citenamefont
  {Dreiser}, \citenamefont {Gambardella}, \citenamefont {Rusponi},\ and\
  \citenamefont {Brune}}]{Donati2014a}%
  \BibitemOpen
  \bibfield  {author} {\bibinfo {author} {\bibfnamefont {F.}~\bibnamefont
  {Donati}}, \bibinfo {author} {\bibfnamefont {A.}~\bibnamefont {Singha}},
  \bibinfo {author} {\bibfnamefont {S.}~\bibnamefont {Stepanow}}, \bibinfo
  {author} {\bibfnamefont {C.}~\bibnamefont {W{\"{a}}ckerlin}}, \bibinfo
  {author} {\bibfnamefont {J.}~\bibnamefont {Dreiser}}, \bibinfo {author}
  {\bibfnamefont {P.}~\bibnamefont {Gambardella}}, \bibinfo {author}
  {\bibfnamefont {S.}~\bibnamefont {Rusponi}},\ and\ \bibinfo {author}
  {\bibfnamefont {H.}~\bibnamefont {Brune}},\ }\href
  {https://doi.org/10.1103/PhysRevLett.113.237201} {\bibfield  {journal}
  {\bibinfo  {journal} {Physical Review Letters}\ }\textbf {\bibinfo {volume}
  {113}},\ \bibinfo {pages} {237201} (\bibinfo {year}
  {2014}{\natexlab{b}})}\BibitemShut {NoStop}%
\bibitem [{\citenamefont {Donati}\ \emph {et~al.}(2016)\citenamefont {Donati},
  \citenamefont {Rusponi}, \citenamefont {Stepanow}, \citenamefont {Wackerlin},
  \citenamefont {Singha}, \citenamefont {Persichetti}, \citenamefont {Baltic},
  \citenamefont {Diller}, \citenamefont {Patthey}, \citenamefont {Fernandes},
  \citenamefont {Dreiser}, \citenamefont {{\v{S}}ljivan{\v{c}}anin},
  \citenamefont {Kummer}, \citenamefont {Nistor}, \citenamefont {Gambardella},\
  and\ \citenamefont {Brune}}]{Donati2016}%
  \BibitemOpen
  \bibfield  {author} {\bibinfo {author} {\bibfnamefont {F.}~\bibnamefont
  {Donati}}, \bibinfo {author} {\bibfnamefont {S.}~\bibnamefont {Rusponi}},
  \bibinfo {author} {\bibfnamefont {S.}~\bibnamefont {Stepanow}}, \bibinfo
  {author} {\bibfnamefont {C.}~\bibnamefont {Wackerlin}}, \bibinfo {author}
  {\bibfnamefont {A.}~\bibnamefont {Singha}}, \bibinfo {author} {\bibfnamefont
  {L.}~\bibnamefont {Persichetti}}, \bibinfo {author} {\bibfnamefont
  {R.}~\bibnamefont {Baltic}}, \bibinfo {author} {\bibfnamefont
  {K.}~\bibnamefont {Diller}}, \bibinfo {author} {\bibfnamefont
  {F.}~\bibnamefont {Patthey}}, \bibinfo {author} {\bibfnamefont
  {E.}~\bibnamefont {Fernandes}}, \bibinfo {author} {\bibfnamefont
  {J.}~\bibnamefont {Dreiser}}, \bibinfo {author} {\bibfnamefont
  {{\v{Z}}.}~\bibnamefont {{\v{S}}ljivan{\v{c}}anin}}, \bibinfo {author}
  {\bibfnamefont {K.}~\bibnamefont {Kummer}}, \bibinfo {author} {\bibfnamefont
  {C.}~\bibnamefont {Nistor}}, \bibinfo {author} {\bibfnamefont
  {P.}~\bibnamefont {Gambardella}},\ and\ \bibinfo {author} {\bibfnamefont
  {H.}~\bibnamefont {Brune}},\ }\href {https://doi.org/10.1126/science.aad9898}
  {\bibfield  {journal} {\bibinfo  {journal} {Science}\ }\textbf {\bibinfo
  {volume} {352}},\ \bibinfo {pages} {318} (\bibinfo {year}
  {2016})}\BibitemShut {NoStop}%
\bibitem [{\citenamefont {Baltic}\ \emph {et~al.}(2018)\citenamefont {Baltic},
  \citenamefont {Donati}, \citenamefont {Singha}, \citenamefont
  {W{\"{a}}ckerlin}, \citenamefont {Dreiser}, \citenamefont {Delley},
  \citenamefont {Pivetta}, \citenamefont {Rusponi},\ and\ \citenamefont
  {Brune}}]{Baltic2018}%
  \BibitemOpen
  \bibfield  {author} {\bibinfo {author} {\bibfnamefont {R.}~\bibnamefont
  {Baltic}}, \bibinfo {author} {\bibfnamefont {F.}~\bibnamefont {Donati}},
  \bibinfo {author} {\bibfnamefont {A.}~\bibnamefont {Singha}}, \bibinfo
  {author} {\bibfnamefont {C.}~\bibnamefont {W{\"{a}}ckerlin}}, \bibinfo
  {author} {\bibfnamefont {J.}~\bibnamefont {Dreiser}}, \bibinfo {author}
  {\bibfnamefont {B.}~\bibnamefont {Delley}}, \bibinfo {author} {\bibfnamefont
  {M.}~\bibnamefont {Pivetta}}, \bibinfo {author} {\bibfnamefont
  {S.}~\bibnamefont {Rusponi}},\ and\ \bibinfo {author} {\bibfnamefont
  {H.}~\bibnamefont {Brune}},\ }\href
  {https://doi.org/10.1103/PhysRevB.98.024412} {\bibfield  {journal} {\bibinfo
  {journal} {Physical Review B}\ }\textbf {\bibinfo {volume} {98}},\ \bibinfo
  {pages} {024412} (\bibinfo {year} {2018})}\BibitemShut {NoStop}%
\bibitem [{\citenamefont {Meier}\ \emph {et~al.}(2008)\citenamefont {Meier},
  \citenamefont {Zhou}, \citenamefont {Wiebe},\ and\ \citenamefont
  {Wiesendanger}}]{Meier2008}%
  \BibitemOpen
  \bibfield  {author} {\bibinfo {author} {\bibfnamefont {F.}~\bibnamefont
  {Meier}}, \bibinfo {author} {\bibfnamefont {L.}~\bibnamefont {Zhou}},
  \bibinfo {author} {\bibfnamefont {J.}~\bibnamefont {Wiebe}},\ and\ \bibinfo
  {author} {\bibfnamefont {R.}~\bibnamefont {Wiesendanger}},\ }\href
  {https://doi.org/10.1126/science.1154415} {\bibfield  {journal} {\bibinfo
  {journal} {Science}\ }\textbf {\bibinfo {volume} {320}},\ \bibinfo {pages}
  {82} (\bibinfo {year} {2008})}\BibitemShut {NoStop}%
\bibitem [{\citenamefont {Loth}\ \emph
  {et~al.}(2010{\natexlab{a}})\citenamefont {Loth}, \citenamefont {von
  Bergmann}, \citenamefont {Ternes}, \citenamefont {Otte}, \citenamefont
  {Lutz},\ and\ \citenamefont {Heinrich}}]{Loth2010}%
  \BibitemOpen
  \bibfield  {author} {\bibinfo {author} {\bibfnamefont {S.}~\bibnamefont
  {Loth}}, \bibinfo {author} {\bibfnamefont {K.}~\bibnamefont {von Bergmann}},
  \bibinfo {author} {\bibfnamefont {M.}~\bibnamefont {Ternes}}, \bibinfo
  {author} {\bibfnamefont {A.~F.}\ \bibnamefont {Otte}}, \bibinfo {author}
  {\bibfnamefont {C.~P.}\ \bibnamefont {Lutz}},\ and\ \bibinfo {author}
  {\bibfnamefont {A.~J.}\ \bibnamefont {Heinrich}},\ }\href
  {https://doi.org/10.1038/nphys1616} {\bibfield  {journal} {\bibinfo
  {journal} {Nature Physics}\ }\textbf {\bibinfo {volume} {6}},\ \bibinfo
  {pages} {340} (\bibinfo {year} {2010}{\natexlab{a}})}\BibitemShut {NoStop}%
\bibitem [{\citenamefont {Loth}\ \emph
  {et~al.}(2010{\natexlab{b}})\citenamefont {Loth}, \citenamefont {Etzkorn},
  \citenamefont {Lutz}, \citenamefont {Eigler},\ and\ \citenamefont
  {Heinrich}}]{Loth2010b}%
  \BibitemOpen
  \bibfield  {author} {\bibinfo {author} {\bibfnamefont {S.}~\bibnamefont
  {Loth}}, \bibinfo {author} {\bibfnamefont {M.}~\bibnamefont {Etzkorn}},
  \bibinfo {author} {\bibfnamefont {C.~P.}\ \bibnamefont {Lutz}}, \bibinfo
  {author} {\bibfnamefont {D.~M.}\ \bibnamefont {Eigler}},\ and\ \bibinfo
  {author} {\bibfnamefont {A.~J.}\ \bibnamefont {Heinrich}},\ }\href
  {https://doi.org/10.1126/science.1191688} {\bibfield  {journal} {\bibinfo
  {journal} {Science}\ }\textbf {\bibinfo {volume} {329}},\ \bibinfo {pages}
  {1628} (\bibinfo {year} {2010}{\natexlab{b}})}\BibitemShut {NoStop}%
\bibitem [{\citenamefont {Paul}\ \emph {et~al.}(2017)\citenamefont {Paul},
  \citenamefont {Yang}, \citenamefont {Baumann}, \citenamefont {Romming},
  \citenamefont {Choi}, \citenamefont {Lutz},\ and\ \citenamefont
  {Heinrich}}]{Paul2017}%
  \BibitemOpen
  \bibfield  {author} {\bibinfo {author} {\bibfnamefont {W.}~\bibnamefont
  {Paul}}, \bibinfo {author} {\bibfnamefont {K.}~\bibnamefont {Yang}}, \bibinfo
  {author} {\bibfnamefont {S.}~\bibnamefont {Baumann}}, \bibinfo {author}
  {\bibfnamefont {N.}~\bibnamefont {Romming}}, \bibinfo {author} {\bibfnamefont
  {T.}~\bibnamefont {Choi}}, \bibinfo {author} {\bibfnamefont {C.~P.}\
  \bibnamefont {Lutz}},\ and\ \bibinfo {author} {\bibfnamefont {A.~J.}\
  \bibnamefont {Heinrich}},\ }\href {https://doi.org/10.1038/nphys3965}
  {\bibfield  {journal} {\bibinfo  {journal} {Nature Physics}\ }\textbf
  {\bibinfo {volume} {13}},\ \bibinfo {pages} {403} (\bibinfo {year}
  {2017})}\BibitemShut {NoStop}%
\bibitem [{\citenamefont {Natterer}\ \emph {et~al.}(2018)\citenamefont
  {Natterer}, \citenamefont {Donati}, \citenamefont {Patthey},\ and\
  \citenamefont {Brune}}]{Natterer2018}%
  \BibitemOpen
  \bibfield  {author} {\bibinfo {author} {\bibfnamefont {F.~D.}\ \bibnamefont
  {Natterer}}, \bibinfo {author} {\bibfnamefont {F.}~\bibnamefont {Donati}},
  \bibinfo {author} {\bibfnamefont {F.}~\bibnamefont {Patthey}},\ and\ \bibinfo
  {author} {\bibfnamefont {H.}~\bibnamefont {Brune}},\ }\href
  {https://doi.org/10.1103/PhysRevLett.121.027201} {\bibfield  {journal}
  {\bibinfo  {journal} {Physical Review Letters}\ }\textbf {\bibinfo {volume}
  {121}},\ \bibinfo {pages} {027201} (\bibinfo {year} {2018})}\BibitemShut
  {NoStop}%
\bibitem [{\citenamefont {Baumann}\ \emph
  {et~al.}(2015{\natexlab{a}})\citenamefont {Baumann}, \citenamefont {Paul},
  \citenamefont {Choi}, \citenamefont {Lutz}, \citenamefont {Ardavan},\ and\
  \citenamefont {Heinrich}}]{Baumann2015b}%
  \BibitemOpen
  \bibfield  {author} {\bibinfo {author} {\bibfnamefont {S.}~\bibnamefont
  {Baumann}}, \bibinfo {author} {\bibfnamefont {W.}~\bibnamefont {Paul}},
  \bibinfo {author} {\bibfnamefont {T.}~\bibnamefont {Choi}}, \bibinfo {author}
  {\bibfnamefont {C.~P.}\ \bibnamefont {Lutz}}, \bibinfo {author}
  {\bibfnamefont {A.}~\bibnamefont {Ardavan}},\ and\ \bibinfo {author}
  {\bibfnamefont {A.~J.}\ \bibnamefont {Heinrich}},\ }\href
  {https://doi.org/10.1126/science.aac8703} {\bibfield  {journal} {\bibinfo
  {journal} {Science}\ }\textbf {\bibinfo {volume} {350}},\ \bibinfo {pages}
  {417} (\bibinfo {year} {2015}{\natexlab{a}})}\BibitemShut {NoStop}%
\bibitem [{\citenamefont {Gatteschi}\ and\ \citenamefont
  {Sessoli}(2003)}]{Gatteschi2003}%
  \BibitemOpen
  \bibfield  {author} {\bibinfo {author} {\bibfnamefont {D.}~\bibnamefont
  {Gatteschi}}\ and\ \bibinfo {author} {\bibfnamefont {R.}~\bibnamefont
  {Sessoli}},\ }\href {https://doi.org/10.1002/anie.200390099} {\bibfield
  {journal} {\bibinfo  {journal} {Angewandte Chemie International Edition}\
  }\textbf {\bibinfo {volume} {42}},\ \bibinfo {pages} {268} (\bibinfo {year}
  {2003})}\BibitemShut {NoStop}%
\bibitem [{\citenamefont {Iba{\~{n}}ez-Azpiroz}\ \emph
  {et~al.}(2017{\natexlab{b}})\citenamefont {Iba{\~{n}}ez-Azpiroz},
  \citenamefont {{dos Santos Dias}}, \citenamefont {Schweflinghaus},
  \citenamefont {Bl{\"{u}}gel},\ and\ \citenamefont
  {Lounis}}]{Ibanez-Azpiroz2017}%
  \BibitemOpen
  \bibfield  {author} {\bibinfo {author} {\bibfnamefont {J.}~\bibnamefont
  {Iba{\~{n}}ez-Azpiroz}}, \bibinfo {author} {\bibfnamefont {M.}~\bibnamefont
  {{dos Santos Dias}}}, \bibinfo {author} {\bibfnamefont {B.}~\bibnamefont
  {Schweflinghaus}}, \bibinfo {author} {\bibfnamefont {S.}~\bibnamefont
  {Bl{\"{u}}gel}},\ and\ \bibinfo {author} {\bibfnamefont {S.}~\bibnamefont
  {Lounis}},\ }\href {https://doi.org/10.1103/PhysRevLett.119.017203}
  {\bibfield  {journal} {\bibinfo  {journal} {Physical Review Letters}\
  }\textbf {\bibinfo {volume} {119}},\ \bibinfo {pages} {017203} (\bibinfo
  {year} {2017}{\natexlab{b}})}\BibitemShut {NoStop}%
\bibitem [{\citenamefont {Shehada}\ \emph {et~al.}(2021)\citenamefont
  {Shehada}, \citenamefont {{dos Santos Dias}}, \citenamefont
  {Guimar{\~{a}}es}, \citenamefont {Abusaa},\ and\ \citenamefont
  {Lounis}}]{Shehada2021}%
  \BibitemOpen
  \bibfield  {author} {\bibinfo {author} {\bibfnamefont {S.}~\bibnamefont
  {Shehada}}, \bibinfo {author} {\bibfnamefont {M.}~\bibnamefont {{dos Santos
  Dias}}}, \bibinfo {author} {\bibfnamefont {F.~S.~M.}\ \bibnamefont
  {Guimar{\~{a}}es}}, \bibinfo {author} {\bibfnamefont {M.}~\bibnamefont
  {Abusaa}},\ and\ \bibinfo {author} {\bibfnamefont {S.}~\bibnamefont
  {Lounis}},\ }\href {https://doi.org/10.1038/s41524-021-00556-y} {\bibfield
  {journal} {\bibinfo  {journal} {npj Computational Materials}\ }\textbf
  {\bibinfo {volume} {7}},\ \bibinfo {pages} {87} (\bibinfo {year}
  {2021})}\BibitemShut {NoStop}%
\bibitem [{\citenamefont {Donati}\ \emph {et~al.}(2020)\citenamefont {Donati},
  \citenamefont {Rusponi}, \citenamefont {Stepanow}, \citenamefont
  {Persichetti}, \citenamefont {Singha}, \citenamefont {Juraschek},
  \citenamefont {W{\"{a}}ckerlin}, \citenamefont {Baltic}, \citenamefont
  {Pivetta}, \citenamefont {Diller}, \citenamefont {Nistor}, \citenamefont
  {Dreiser}, \citenamefont {Kummer}, \citenamefont {Velez-Fort}, \citenamefont
  {Spaldin}, \citenamefont {Brune},\ and\ \citenamefont
  {Gambardella}}]{Donati2020}%
  \BibitemOpen
  \bibfield  {author} {\bibinfo {author} {\bibfnamefont {F.}~\bibnamefont
  {Donati}}, \bibinfo {author} {\bibfnamefont {S.}~\bibnamefont {Rusponi}},
  \bibinfo {author} {\bibfnamefont {S.}~\bibnamefont {Stepanow}}, \bibinfo
  {author} {\bibfnamefont {L.}~\bibnamefont {Persichetti}}, \bibinfo {author}
  {\bibfnamefont {A.}~\bibnamefont {Singha}}, \bibinfo {author} {\bibfnamefont
  {D.~M.}\ \bibnamefont {Juraschek}}, \bibinfo {author} {\bibfnamefont
  {C.}~\bibnamefont {W{\"{a}}ckerlin}}, \bibinfo {author} {\bibfnamefont
  {R.}~\bibnamefont {Baltic}}, \bibinfo {author} {\bibfnamefont
  {M.}~\bibnamefont {Pivetta}}, \bibinfo {author} {\bibfnamefont
  {K.}~\bibnamefont {Diller}}, \bibinfo {author} {\bibfnamefont
  {C.}~\bibnamefont {Nistor}}, \bibinfo {author} {\bibfnamefont
  {J.}~\bibnamefont {Dreiser}}, \bibinfo {author} {\bibfnamefont
  {K.}~\bibnamefont {Kummer}}, \bibinfo {author} {\bibfnamefont
  {E.}~\bibnamefont {Velez-Fort}}, \bibinfo {author} {\bibfnamefont {N.~A.}\
  \bibnamefont {Spaldin}}, \bibinfo {author} {\bibfnamefont {H.}~\bibnamefont
  {Brune}},\ and\ \bibinfo {author} {\bibfnamefont {P.}~\bibnamefont
  {Gambardella}},\ }\href {https://doi.org/10.1103/PhysRevLett.124.077204}
  {\bibfield  {journal} {\bibinfo  {journal} {Physical Review Letters}\
  }\textbf {\bibinfo {volume} {124}},\ \bibinfo {pages} {077204} (\bibinfo
  {year} {2020})}\BibitemShut {NoStop}%
\bibitem [{\citenamefont {Giustino}(2017)}]{Giustino2017}%
  \BibitemOpen
  \bibfield  {author} {\bibinfo {author} {\bibfnamefont {F.}~\bibnamefont
  {Giustino}},\ }\href {https://doi.org/10.1103/RevModPhys.89.015003}
  {\bibfield  {journal} {\bibinfo  {journal} {Reviews of Modern Physics}\
  }\textbf {\bibinfo {volume} {89}},\ \bibinfo {pages} {015003} (\bibinfo
  {year} {2017})}\BibitemShut {NoStop}%
\bibitem [{\citenamefont {Roychoudhury}\ and\ \citenamefont
  {Sanvito}(2018)}]{Roychoudhury2018}%
  \BibitemOpen
  \bibfield  {author} {\bibinfo {author} {\bibfnamefont {S.}~\bibnamefont
  {Roychoudhury}}\ and\ \bibinfo {author} {\bibfnamefont {S.}~\bibnamefont
  {Sanvito}},\ }\href {https://doi.org/10.1103/PhysRevB.98.125204} {\bibfield
  {journal} {\bibinfo  {journal} {Physical Review B}\ }\textbf {\bibinfo
  {volume} {98}},\ \bibinfo {pages} {125204} (\bibinfo {year}
  {2018})}\BibitemShut {NoStop}%
\bibitem [{\citenamefont {Lunghi}\ and\ \citenamefont
  {Sanvito}(2019)}]{Lunghi2019}%
  \BibitemOpen
  \bibfield  {author} {\bibinfo {author} {\bibfnamefont {A.}~\bibnamefont
  {Lunghi}}\ and\ \bibinfo {author} {\bibfnamefont {S.}~\bibnamefont
  {Sanvito}},\ }\href {https://doi.org/10.1126/sciadv.aax7163} {\bibfield
  {journal} {\bibinfo  {journal} {Science Advances}\ }\textbf {\bibinfo
  {volume} {5}},\ \bibinfo {pages} {eaax7163} (\bibinfo {year}
  {2019})}\BibitemShut {NoStop}%
\bibitem [{\citenamefont {Hohenberg}\ and\ \citenamefont
  {Kohn}(1964)}]{Hohenberg1964}%
  \BibitemOpen
  \bibfield  {author} {\bibinfo {author} {\bibfnamefont {P.}~\bibnamefont
  {Hohenberg}}\ and\ \bibinfo {author} {\bibfnamefont {W.}~\bibnamefont
  {Kohn}},\ }\href {https://doi.org/10.1103/PhysRev.136.B864} {\bibfield
  {journal} {\bibinfo  {journal} {Physical Review}\ }\textbf {\bibinfo {volume}
  {136}},\ \bibinfo {pages} {B864} (\bibinfo {year} {1964})}\BibitemShut
  {NoStop}%
\bibitem [{\citenamefont {Kohn}\ and\ \citenamefont {Sham}(1965)}]{Kohn1965}%
  \BibitemOpen
  \bibfield  {author} {\bibinfo {author} {\bibfnamefont {W.}~\bibnamefont
  {Kohn}}\ and\ \bibinfo {author} {\bibfnamefont {L.~J.}\ \bibnamefont
  {Sham}},\ }\href {https://doi.org/10.1103/PhysRev.140.A1133} {\bibfield
  {journal} {\bibinfo  {journal} {Physical Review}\ }\textbf {\bibinfo {volume}
  {140}},\ \bibinfo {pages} {A1133} (\bibinfo {year} {1965})}\BibitemShut
  {NoStop}%
\bibitem [{\citenamefont {Soler}\ \emph {et~al.}(2002)\citenamefont {Soler},
  \citenamefont {Artacho}, \citenamefont {Gale}, \citenamefont {Garc{\'{i}}a},
  \citenamefont {Junquera}, \citenamefont {Ordej{\'{o}}n},\ and\ \citenamefont
  {S{\'{a}}nchez-Portal}}]{Soler2002}%
  \BibitemOpen
  \bibfield  {author} {\bibinfo {author} {\bibfnamefont {J.~M.}\ \bibnamefont
  {Soler}}, \bibinfo {author} {\bibfnamefont {E.}~\bibnamefont {Artacho}},
  \bibinfo {author} {\bibfnamefont {J.~D.}\ \bibnamefont {Gale}}, \bibinfo
  {author} {\bibfnamefont {A.}~\bibnamefont {Garc{\'{i}}a}}, \bibinfo {author}
  {\bibfnamefont {J.}~\bibnamefont {Junquera}}, \bibinfo {author}
  {\bibfnamefont {P.}~\bibnamefont {Ordej{\'{o}}n}},\ and\ \bibinfo {author}
  {\bibfnamefont {D.}~\bibnamefont {S{\'{a}}nchez-Portal}},\ }\href
  {https://doi.org/10.1088/0953-8984/14/11/302} {\bibfield  {journal} {\bibinfo
   {journal} {Journal of Physics: Condensed Matter}\ }\textbf {\bibinfo
  {volume} {14}},\ \bibinfo {pages} {2745} (\bibinfo {year}
  {2002})}\BibitemShut {NoStop}%
\bibitem [{\citenamefont {Pickett}(1989)}]{Pickett1989}%
  \BibitemOpen
  \bibfield  {author} {\bibinfo {author} {\bibfnamefont {W.~E.}\ \bibnamefont
  {Pickett}},\ }\href {https://doi.org/10.1016/0167-7977(89)90002-6} {\bibfield
   {journal} {\bibinfo  {journal} {Computer Physics Reports}\ }\textbf
  {\bibinfo {volume} {9}},\ \bibinfo {pages} {115} (\bibinfo {year}
  {1989})}\BibitemShut {NoStop}%
\bibitem [{\citenamefont {Kleinman}\ and\ \citenamefont
  {Bylander}(1982)}]{Kleinman1982}%
  \BibitemOpen
  \bibfield  {author} {\bibinfo {author} {\bibfnamefont {L.}~\bibnamefont
  {Kleinman}}\ and\ \bibinfo {author} {\bibfnamefont {D.~M.}\ \bibnamefont
  {Bylander}},\ }\href {https://doi.org/10.1103/PhysRevLett.48.1425} {\bibfield
   {journal} {\bibinfo  {journal} {Physical Review Letters}\ }\textbf {\bibinfo
  {volume} {48}},\ \bibinfo {pages} {1425} (\bibinfo {year}
  {1982})}\BibitemShut {NoStop}%
\bibitem [{\citenamefont {Hamann}\ \emph {et~al.}(1979)\citenamefont {Hamann},
  \citenamefont {Schl{\"{u}}ter},\ and\ \citenamefont {Chiang}}]{Hamann1979}%
  \BibitemOpen
  \bibfield  {author} {\bibinfo {author} {\bibfnamefont {D.~R.}\ \bibnamefont
  {Hamann}}, \bibinfo {author} {\bibfnamefont {M.}~\bibnamefont
  {Schl{\"{u}}ter}},\ and\ \bibinfo {author} {\bibfnamefont {C.}~\bibnamefont
  {Chiang}},\ }\href {https://doi.org/10.1103/PhysRevLett.43.1494} {\bibfield
  {journal} {\bibinfo  {journal} {Physical Review Letters}\ }\textbf {\bibinfo
  {volume} {43}},\ \bibinfo {pages} {1494} (\bibinfo {year}
  {1979})}\BibitemShut {NoStop}%
\bibitem [{\citenamefont {Perdew}\ \emph {et~al.}(1996)\citenamefont {Perdew},
  \citenamefont {Burke},\ and\ \citenamefont {Ernzerhof}}]{Perdew1996}%
  \BibitemOpen
  \bibfield  {author} {\bibinfo {author} {\bibfnamefont {J.~P.}\ \bibnamefont
  {Perdew}}, \bibinfo {author} {\bibfnamefont {K.}~\bibnamefont {Burke}},\ and\
  \bibinfo {author} {\bibfnamefont {M.}~\bibnamefont {Ernzerhof}},\ }\href
  {https://doi.org/10.1103/PhysRevLett.77.3865} {\bibfield  {journal} {\bibinfo
   {journal} {Physical Review Letters}\ }\textbf {\bibinfo {volume} {77}},\
  \bibinfo {pages} {3865} (\bibinfo {year} {1996})}\BibitemShut {NoStop}%
\bibitem [{\citenamefont {Monkhorst}\ and\ \citenamefont
  {Pack}(1976)}]{Monkhorst1976}%
  \BibitemOpen
  \bibfield  {author} {\bibinfo {author} {\bibfnamefont {H.~J.}\ \bibnamefont
  {Monkhorst}}\ and\ \bibinfo {author} {\bibfnamefont {J.~D.}\ \bibnamefont
  {Pack}},\ }\href {https://doi.org/10.1103/PhysRevB.13.5188} {\bibfield
  {journal} {\bibinfo  {journal} {Physical Review B}\ }\textbf {\bibinfo
  {volume} {13}},\ \bibinfo {pages} {5188} (\bibinfo {year}
  {1976})}\BibitemShut {NoStop}%
\bibitem [{\citenamefont {Baroni}\ \emph {et~al.}(2001)\citenamefont {Baroni},
  \citenamefont {de~Gironcoli}, \citenamefont {{Dal Corso}},\ and\
  \citenamefont {Giannozzi}}]{Baroni2001}%
  \BibitemOpen
  \bibfield  {author} {\bibinfo {author} {\bibfnamefont {S.}~\bibnamefont
  {Baroni}}, \bibinfo {author} {\bibfnamefont {S.}~\bibnamefont
  {de~Gironcoli}}, \bibinfo {author} {\bibfnamefont {A.}~\bibnamefont {{Dal
  Corso}}},\ and\ \bibinfo {author} {\bibfnamefont {P.}~\bibnamefont
  {Giannozzi}},\ }\href {https://doi.org/10.1103/RevModPhys.73.515} {\bibfield
  {journal} {\bibinfo  {journal} {Reviews of Modern Physics}\ }\textbf
  {\bibinfo {volume} {73}},\ \bibinfo {pages} {515} (\bibinfo {year}
  {2001})}\BibitemShut {NoStop}%
\bibitem [{\citenamefont {Frank}\ \emph {et~al.}(1995)\citenamefont {Frank},
  \citenamefont {Els{\"{a}}sser},\ and\ \citenamefont
  {F{\"{a}}hnle}}]{Frank1995}%
  \BibitemOpen
  \bibfield  {author} {\bibinfo {author} {\bibfnamefont {W.}~\bibnamefont
  {Frank}}, \bibinfo {author} {\bibfnamefont {C.}~\bibnamefont
  {Els{\"{a}}sser}},\ and\ \bibinfo {author} {\bibfnamefont {M.}~\bibnamefont
  {F{\"{a}}hnle}},\ }\href {https://doi.org/10.1103/PhysRevLett.74.1791}
  {\bibfield  {journal} {\bibinfo  {journal} {Physical Review Letters}\
  }\textbf {\bibinfo {volume} {74}},\ \bibinfo {pages} {1791} (\bibinfo {year}
  {1995})}\BibitemShut {NoStop}%
\bibitem [{\citenamefont {Parlinski}\ \emph {et~al.}(1997)\citenamefont
  {Parlinski}, \citenamefont {Li},\ and\ \citenamefont
  {Kawazoe}}]{Parlinski1997}%
  \BibitemOpen
  \bibfield  {author} {\bibinfo {author} {\bibfnamefont {K.}~\bibnamefont
  {Parlinski}}, \bibinfo {author} {\bibfnamefont {Z.~Q.}\ \bibnamefont {Li}},\
  and\ \bibinfo {author} {\bibfnamefont {Y.}~\bibnamefont {Kawazoe}},\ }\href
  {https://doi.org/10.1103/PhysRevLett.78.4063} {\bibfield  {journal} {\bibinfo
   {journal} {Physical Review Letters}\ }\textbf {\bibinfo {volume} {78}},\
  \bibinfo {pages} {4063} (\bibinfo {year} {1997})}\BibitemShut {NoStop}%
\bibitem [{Note1()}]{Note1}%
  \BibitemOpen
  \bibinfo {note} {16 for the substrate + 3 for the iron + 3 for the oxygen
  below it.}\BibitemShut {Stop}%
\bibitem [{\citenamefont {Chaput}\ \emph {et~al.}(2011)\citenamefont {Chaput},
  \citenamefont {Togo}, \citenamefont {Tanaka},\ and\ \citenamefont
  {Hug}}]{Chaput2011}%
  \BibitemOpen
  \bibfield  {author} {\bibinfo {author} {\bibfnamefont {L.}~\bibnamefont
  {Chaput}}, \bibinfo {author} {\bibfnamefont {A.}~\bibnamefont {Togo}},
  \bibinfo {author} {\bibfnamefont {I.}~\bibnamefont {Tanaka}},\ and\ \bibinfo
  {author} {\bibfnamefont {G.}~\bibnamefont {Hug}},\ }\href
  {https://doi.org/10.1103/PhysRevB.84.094302} {\bibfield  {journal} {\bibinfo
  {journal} {Physical Review B}\ }\textbf {\bibinfo {volume} {84}},\ \bibinfo
  {pages} {094302} (\bibinfo {year} {2011})}\BibitemShut {NoStop}%
\bibitem [{\citenamefont {Garcia-Goiricelaya}\ \emph
  {et~al.}(2018)\citenamefont {Garcia-Goiricelaya}, \citenamefont {{G.
  Gurtubay}},\ and\ \citenamefont {Eiguren}}]{Garcia-Goiricelaya2018}%
  \BibitemOpen
  \bibfield  {author} {\bibinfo {author} {\bibfnamefont {P.}~\bibnamefont
  {Garcia-Goiricelaya}}, \bibinfo {author} {\bibfnamefont {I.}~\bibnamefont
  {{G. Gurtubay}}},\ and\ \bibinfo {author} {\bibfnamefont {A.}~\bibnamefont
  {Eiguren}},\ }\href {https://doi.org/10.1103/PhysRevB.97.201405} {\bibfield
  {journal} {\bibinfo  {journal} {Physical Review B}\ }\textbf {\bibinfo
  {volume} {97}},\ \bibinfo {pages} {201405(R)} (\bibinfo {year}
  {2018})}\BibitemShut {NoStop}%
\bibitem [{\citenamefont {Garcia-Goiricelaya}\ \emph
  {et~al.}(2019)\citenamefont {Garcia-Goiricelaya}, \citenamefont
  {Lafuente-Bartolome}, \citenamefont {Gurtubay},\ and\ \citenamefont
  {Eiguren}}]{Garcia-Goiricelaya2019}%
  \BibitemOpen
  \bibfield  {author} {\bibinfo {author} {\bibfnamefont {P.}~\bibnamefont
  {Garcia-Goiricelaya}}, \bibinfo {author} {\bibfnamefont {J.}~\bibnamefont
  {Lafuente-Bartolome}}, \bibinfo {author} {\bibfnamefont {I.~G.}\ \bibnamefont
  {Gurtubay}},\ and\ \bibinfo {author} {\bibfnamefont {A.}~\bibnamefont
  {Eiguren}},\ }\href {https://doi.org/10.1038/s42005-019-0182-0} {\bibfield
  {journal} {\bibinfo  {journal} {Communications Physics}\ }\textbf {\bibinfo
  {volume} {2}},\ \bibinfo {pages} {81} (\bibinfo {year} {2019})}\BibitemShut
  {NoStop}%
\bibitem [{\citenamefont {Garcia-Goiricelaya}\ \emph
  {et~al.}(2020)\citenamefont {Garcia-Goiricelaya}, \citenamefont
  {Lafuente-Bartolome}, \citenamefont {{G. Gurtubay}},\ and\ \citenamefont
  {Eiguren}}]{Garcia-Goiricelaya2020}%
  \BibitemOpen
  \bibfield  {author} {\bibinfo {author} {\bibfnamefont {P.}~\bibnamefont
  {Garcia-Goiricelaya}}, \bibinfo {author} {\bibfnamefont {J.}~\bibnamefont
  {Lafuente-Bartolome}}, \bibinfo {author} {\bibfnamefont {I.}~\bibnamefont
  {{G. Gurtubay}}},\ and\ \bibinfo {author} {\bibfnamefont {A.}~\bibnamefont
  {Eiguren}},\ }\href {https://doi.org/10.1103/PhysRevB.101.054304} {\bibfield
  {journal} {\bibinfo  {journal} {Physical Review B}\ }\textbf {\bibinfo
  {volume} {101}},\ \bibinfo {pages} {054304} (\bibinfo {year}
  {2020})}\BibitemShut {NoStop}%
\bibitem [{\citenamefont {Lafuente-Bartolome}\ \emph
  {et~al.}(2020{\natexlab{a}})\citenamefont {Lafuente-Bartolome}, \citenamefont
  {{G. Gurtubay}},\ and\ \citenamefont {Eiguren}}]{Lafuente-Bartolome2020}%
  \BibitemOpen
  \bibfield  {author} {\bibinfo {author} {\bibfnamefont {J.}~\bibnamefont
  {Lafuente-Bartolome}}, \bibinfo {author} {\bibfnamefont {I.}~\bibnamefont
  {{G. Gurtubay}}},\ and\ \bibinfo {author} {\bibfnamefont {A.}~\bibnamefont
  {Eiguren}},\ }\href {https://doi.org/10.1103/PhysRevB.102.165113} {\bibfield
  {journal} {\bibinfo  {journal} {Physical Review B}\ }\textbf {\bibinfo
  {volume} {102}},\ \bibinfo {pages} {165113} (\bibinfo {year}
  {2020}{\natexlab{a}})}\BibitemShut {NoStop}%
\bibitem [{\citenamefont {Lafuente-Bartolome}\ \emph
  {et~al.}(2020{\natexlab{b}})\citenamefont {Lafuente-Bartolome}, \citenamefont
  {Gurtubay},\ and\ \citenamefont {Eiguren}}]{Lafuente-Bartolome2020a}%
  \BibitemOpen
  \bibfield  {author} {\bibinfo {author} {\bibfnamefont {J.}~\bibnamefont
  {Lafuente-Bartolome}}, \bibinfo {author} {\bibfnamefont {I.~G.}\ \bibnamefont
  {Gurtubay}},\ and\ \bibinfo {author} {\bibfnamefont {A.}~\bibnamefont
  {Eiguren}},\ }\href {https://doi.org/10.1103/PhysRevB.102.161107} {\bibfield
  {journal} {\bibinfo  {journal} {Physical Review B}\ }\textbf {\bibinfo
  {volume} {102}},\ \bibinfo {pages} {161107(R)} (\bibinfo {year}
  {2020}{\natexlab{b}})}\BibitemShut {NoStop}%
\bibitem [{\citenamefont {Kolb}\ \emph {et~al.}(1981)\citenamefont {Kolb},
  \citenamefont {Boeck}, \citenamefont {Ho},\ and\ \citenamefont
  {Liu}}]{Kolb1981}%
  \BibitemOpen
  \bibfield  {author} {\bibinfo {author} {\bibfnamefont {D.~M.}\ \bibnamefont
  {Kolb}}, \bibinfo {author} {\bibfnamefont {W.}~\bibnamefont {Boeck}},
  \bibinfo {author} {\bibfnamefont {K.-M.}\ \bibnamefont {Ho}},\ and\ \bibinfo
  {author} {\bibfnamefont {S.~H.}\ \bibnamefont {Liu}},\ }\href
  {https://doi.org/10.1103/PhysRevLett.47.1921} {\bibfield  {journal} {\bibinfo
   {journal} {Physical Review Letters}\ }\textbf {\bibinfo {volume} {47}},\
  \bibinfo {pages} {1921} (\bibinfo {year} {1981})}\BibitemShut {NoStop}%
\bibitem [{\citenamefont {Reihl}\ \emph {et~al.}(1984)\citenamefont {Reihl},
  \citenamefont {Frank},\ and\ \citenamefont {Schlittler}}]{Reihl1984}%
  \BibitemOpen
  \bibfield  {author} {\bibinfo {author} {\bibfnamefont {B.}~\bibnamefont
  {Reihl}}, \bibinfo {author} {\bibfnamefont {K.~H.}\ \bibnamefont {Frank}},\
  and\ \bibinfo {author} {\bibfnamefont {R.~R.}\ \bibnamefont {Schlittler}},\
  }\href {https://doi.org/10.1103/PhysRevB.30.7328} {\bibfield  {journal}
  {\bibinfo  {journal} {Physical Review B}\ }\textbf {\bibinfo {volume} {30}},\
  \bibinfo {pages} {7328} (\bibinfo {year} {1984})}\BibitemShut {NoStop}%
\bibitem [{\citenamefont {Altmann}\ \emph {et~al.}(1986)\citenamefont
  {Altmann}, \citenamefont {Dose},\ and\ \citenamefont
  {Goldmann}}]{Altmann1986}%
  \BibitemOpen
  \bibfield  {author} {\bibinfo {author} {\bibfnamefont {W.}~\bibnamefont
  {Altmann}}, \bibinfo {author} {\bibfnamefont {V.}~\bibnamefont {Dose}},\ and\
  \bibinfo {author} {\bibfnamefont {A.}~\bibnamefont {Goldmann}},\ }\href
  {https://doi.org/10.1007/BF01303842} {\bibfield  {journal} {\bibinfo
  {journal} {Zeitschrift f{\"{u}}r Physik B Condensed Matterr Physik B
  Condensed Matter}\ }\textbf {\bibinfo {volume} {65}},\ \bibinfo {pages} {171}
  (\bibinfo {year} {1986})}\BibitemShut {NoStop}%
\bibitem [{\citenamefont {Reihl}\ and\ \citenamefont
  {Nicholls}(1987)}]{Reihl1987}%
  \BibitemOpen
  \bibfield  {author} {\bibinfo {author} {\bibfnamefont {B.}~\bibnamefont
  {Reihl}}\ and\ \bibinfo {author} {\bibfnamefont {J.~M.}\ \bibnamefont
  {Nicholls}},\ }\href {https://doi.org/10.1007/BF01303985} {\bibfield
  {journal} {\bibinfo  {journal} {Zeitschrift f{\"{u}}r Physik B Condensed
  Matter}\ }\textbf {\bibinfo {volume} {67}},\ \bibinfo {pages} {221} (\bibinfo
  {year} {1987})}\BibitemShut {NoStop}%
\bibitem [{\citenamefont {Erschbaumer}\ \emph {et~al.}(1991)\citenamefont
  {Erschbaumer}, \citenamefont {Freeman}, \citenamefont {Fu},\ and\
  \citenamefont {Podloucky}}]{Erschbaumer1991}%
  \BibitemOpen
  \bibfield  {author} {\bibinfo {author} {\bibfnamefont {H.}~\bibnamefont
  {Erschbaumer}}, \bibinfo {author} {\bibfnamefont {A.}~\bibnamefont
  {Freeman}}, \bibinfo {author} {\bibfnamefont {C.}~\bibnamefont {Fu}},\ and\
  \bibinfo {author} {\bibfnamefont {R.}~\bibnamefont {Podloucky}},\ }\href
  {https://doi.org/10.1016/0039-6028(91)90369-4} {\bibfield  {journal}
  {\bibinfo  {journal} {Surface Science}\ }\textbf {\bibinfo {volume} {243}},\
  \bibinfo {pages} {317} (\bibinfo {year} {1991})}\BibitemShut {NoStop}%
\bibitem [{\citenamefont {Savio}\ \emph {et~al.}(2001)\citenamefont {Savio},
  \citenamefont {Vattuone}, \citenamefont {Rocca}, \citenamefont {{De Renzi}},
  \citenamefont {Gardonio}, \citenamefont {Mariani}, \citenamefont {del
  Pennino}, \citenamefont {Cipriani}, \citenamefont {{Dal Corso}},\ and\
  \citenamefont {Baroni}}]{Savio2001}%
  \BibitemOpen
  \bibfield  {author} {\bibinfo {author} {\bibfnamefont {L.}~\bibnamefont
  {Savio}}, \bibinfo {author} {\bibfnamefont {L.}~\bibnamefont {Vattuone}},
  \bibinfo {author} {\bibfnamefont {M.}~\bibnamefont {Rocca}}, \bibinfo
  {author} {\bibfnamefont {V.}~\bibnamefont {{De Renzi}}}, \bibinfo {author}
  {\bibfnamefont {S.}~\bibnamefont {Gardonio}}, \bibinfo {author}
  {\bibfnamefont {C.}~\bibnamefont {Mariani}}, \bibinfo {author} {\bibfnamefont
  {U.}~\bibnamefont {del Pennino}}, \bibinfo {author} {\bibfnamefont
  {G.}~\bibnamefont {Cipriani}}, \bibinfo {author} {\bibfnamefont
  {A.}~\bibnamefont {{Dal Corso}}},\ and\ \bibinfo {author} {\bibfnamefont
  {S.}~\bibnamefont {Baroni}},\ }\href
  {https://doi.org/10.1016/S0039-6028(01)01026-3} {\bibfield  {journal}
  {\bibinfo  {journal} {Surface Science}\ }\textbf {\bibinfo {volume} {486}},\
  \bibinfo {pages} {65} (\bibinfo {year} {2001})}\BibitemShut {NoStop}%
\bibitem [{\citenamefont {Heid}\ and\ \citenamefont {Bohnen}(2003)}]{Heid2003}%
  \BibitemOpen
  \bibfield  {author} {\bibinfo {author} {\bibfnamefont {R.}~\bibnamefont
  {Heid}}\ and\ \bibinfo {author} {\bibfnamefont {K.-P.}\ \bibnamefont
  {Bohnen}},\ }\href {https://doi.org/10.1016/j.physrep.2003.07.003} {\bibfield
   {journal} {\bibinfo  {journal} {Physics Reports}\ }\textbf {\bibinfo
  {volume} {387}},\ \bibinfo {pages} {151} (\bibinfo {year}
  {2003})}\BibitemShut {NoStop}%
\bibitem [{\citenamefont {Baumann}\ \emph
  {et~al.}(2015{\natexlab{b}})\citenamefont {Baumann}, \citenamefont {Donati},
  \citenamefont {Stepanow}, \citenamefont {Rusponi}, \citenamefont {Paul},
  \citenamefont {Gangopadhyay}, \citenamefont {Rau}, \citenamefont {Pacchioni},
  \citenamefont {Gragnaniello}, \citenamefont {Pivetta}, \citenamefont
  {Dreiser}, \citenamefont {Piamonteze}, \citenamefont {Lutz}, \citenamefont
  {Macfarlane}, \citenamefont {Jones}, \citenamefont {Gambardella},
  \citenamefont {Heinrich},\ and\ \citenamefont {Brune}}]{Baumann2015}%
  \BibitemOpen
  \bibfield  {author} {\bibinfo {author} {\bibfnamefont {S.}~\bibnamefont
  {Baumann}}, \bibinfo {author} {\bibfnamefont {F.}~\bibnamefont {Donati}},
  \bibinfo {author} {\bibfnamefont {S.}~\bibnamefont {Stepanow}}, \bibinfo
  {author} {\bibfnamefont {S.}~\bibnamefont {Rusponi}}, \bibinfo {author}
  {\bibfnamefont {W.}~\bibnamefont {Paul}}, \bibinfo {author} {\bibfnamefont
  {S.}~\bibnamefont {Gangopadhyay}}, \bibinfo {author} {\bibfnamefont {I.~G.}\
  \bibnamefont {Rau}}, \bibinfo {author} {\bibfnamefont {G.~E.}\ \bibnamefont
  {Pacchioni}}, \bibinfo {author} {\bibfnamefont {L.}~\bibnamefont
  {Gragnaniello}}, \bibinfo {author} {\bibfnamefont {M.}~\bibnamefont
  {Pivetta}}, \bibinfo {author} {\bibfnamefont {J.}~\bibnamefont {Dreiser}},
  \bibinfo {author} {\bibfnamefont {C.}~\bibnamefont {Piamonteze}}, \bibinfo
  {author} {\bibfnamefont {C.~P.}\ \bibnamefont {Lutz}}, \bibinfo {author}
  {\bibfnamefont {R.~M.}\ \bibnamefont {Macfarlane}}, \bibinfo {author}
  {\bibfnamefont {B.~A.}\ \bibnamefont {Jones}}, \bibinfo {author}
  {\bibfnamefont {P.}~\bibnamefont {Gambardella}}, \bibinfo {author}
  {\bibfnamefont {A.~J.}\ \bibnamefont {Heinrich}},\ and\ \bibinfo {author}
  {\bibfnamefont {H.}~\bibnamefont {Brune}},\ }\href
  {https://doi.org/10.1103/PhysRevLett.115.237202} {\bibfield  {journal}
  {\bibinfo  {journal} {Physical Review Letters}\ }\textbf {\bibinfo {volume}
  {115}},\ \bibinfo {pages} {237202} (\bibinfo {year}
  {2015}{\natexlab{b}})}\BibitemShut {NoStop}%
\bibitem [{\citenamefont {Grimvall}(1983)}]{Grimvall1981}%
  \BibitemOpen
  \bibfield  {author} {\bibinfo {author} {\bibfnamefont {G.}~\bibnamefont
  {Grimvall}},\ }\href {https://doi.org/10.1002/bbpc.19830870521} {\emph
  {\bibinfo {title} {{The Electron-Phonon Interaction in Metals}}}},\ Series of
  monographs on selected topics in solid state physics\ (\bibinfo  {publisher}
  {North Holland Publishing Company},\ \bibinfo {address} {Amsterdam, New York,
  Oxford},\ \bibinfo {year} {1983})\BibitemShut {NoStop}%
\bibitem [{\citenamefont {Mahan}(2000)}]{Mahan2000}%
  \BibitemOpen
  \bibfield  {author} {\bibinfo {author} {\bibfnamefont {G.~D.}\ \bibnamefont
  {Mahan}},\ }\href {https://doi.org/10.1007/978-1-4757-5714-9} {\emph
  {\bibinfo {title} {{Many-Particle Physics}}}},\ \bibinfo {edition} {3rd}\
  ed.\ (\bibinfo  {publisher} {Springer US},\ \bibinfo {address} {Boston, MA},\
  \bibinfo {year} {2000})\BibitemShut {NoStop}%
\bibitem [{\citenamefont {Sklyadneva}\ \emph {et~al.}(2012)\citenamefont
  {Sklyadneva}, \citenamefont {Heid}, \citenamefont {Echenique}, \citenamefont
  {Bohnen},\ and\ \citenamefont {Chulkov}}]{Sklyadneva2012}%
  \BibitemOpen
  \bibfield  {author} {\bibinfo {author} {\bibfnamefont {I.~Y.}\ \bibnamefont
  {Sklyadneva}}, \bibinfo {author} {\bibfnamefont {R.}~\bibnamefont {Heid}},
  \bibinfo {author} {\bibfnamefont {P.~M.}\ \bibnamefont {Echenique}}, \bibinfo
  {author} {\bibfnamefont {K.-B.}\ \bibnamefont {Bohnen}},\ and\ \bibinfo
  {author} {\bibfnamefont {E.~V.}\ \bibnamefont {Chulkov}},\ }\href
  {https://doi.org/10.1103/PhysRevB.85.155115} {\bibfield  {journal} {\bibinfo
  {journal} {Physical Review B}\ }\textbf {\bibinfo {volume} {85}},\ \bibinfo
  {pages} {155115} (\bibinfo {year} {2012})}\BibitemShut {NoStop}%
\bibitem [{\citenamefont {Hofmann}\ \emph {et~al.}(2009)\citenamefont
  {Hofmann}, \citenamefont {Sklyadneva}, \citenamefont {Rienks},\ and\
  \citenamefont {Chulkov}}]{Hofmann2009}%
  \BibitemOpen
  \bibfield  {author} {\bibinfo {author} {\bibfnamefont {P.}~\bibnamefont
  {Hofmann}}, \bibinfo {author} {\bibfnamefont {I.~Y.}\ \bibnamefont
  {Sklyadneva}}, \bibinfo {author} {\bibfnamefont {E.~D.~L.}\ \bibnamefont
  {Rienks}},\ and\ \bibinfo {author} {\bibfnamefont {E.~V.}\ \bibnamefont
  {Chulkov}},\ }\href {https://doi.org/10.1088/1367-2630/11/12/125005}
  {\bibfield  {journal} {\bibinfo  {journal} {New Journal of Physics}\ }\textbf
  {\bibinfo {volume} {11}},\ \bibinfo {pages} {125005} (\bibinfo {year}
  {2009})}\BibitemShut {NoStop}%
\bibitem [{\citenamefont {Lunghi}\ \emph {et~al.}(2017)\citenamefont {Lunghi},
  \citenamefont {Totti}, \citenamefont {Sessoli},\ and\ \citenamefont
  {Sanvito}}]{Lunghi2017}%
  \BibitemOpen
  \bibfield  {author} {\bibinfo {author} {\bibfnamefont {A.}~\bibnamefont
  {Lunghi}}, \bibinfo {author} {\bibfnamefont {F.}~\bibnamefont {Totti}},
  \bibinfo {author} {\bibfnamefont {R.}~\bibnamefont {Sessoli}},\ and\ \bibinfo
  {author} {\bibfnamefont {S.}~\bibnamefont {Sanvito}},\ }\href
  {https://doi.org/10.1038/ncomms14620} {\bibfield  {journal} {\bibinfo
  {journal} {Nature Communications}\ }\textbf {\bibinfo {volume} {8}},\
  \bibinfo {pages} {14620} (\bibinfo {year} {2017})}\BibitemShut {NoStop}%
\end{thebibliography}%

\end{document}